\documentclass[fleqn,usenatbib,useAMS]{mnras}

\usepackage{newtxtext,newtxmath}

\usepackage[T1]{fontenc}

\DeclareRobustCommand{\VAN}[3]{#2}
\let\VANthebibliography\thebibliography
\def\thebibliography{\DeclareRobustCommand{\VAN}[3]{##3}\VANthebibliography}

\usepackage{graphicx}	
\usepackage{amsmath}	
\usepackage{xcolor}
\usepackage{multicol}

\title[PRIMAger: Overcoming Confusion]{Overcoming Confusion Noise with Hyperspectral Imaging from PRIMAger}

\author[]{
J. M. S. Donnellan,$^{1}$\thanks{E-mail: j.donnellan@sussex.ac.uk}
S. J. Oliver,$^{1}$
M. B\'ethermin,$^{2,3}$
L. Bing,$^{1}$
A. Bolatto,$^{4}$
C. M. Bradford,$^{5,6}$
\newauthor{D. Burgarella,$^{3}$}
L. Ciesla,$^{3}$
J. Glenn,$^{7}$
A. Pope,$^{8}$
S. Serjeant,$^{9}$
R. Shirley,$^{10}$
J. D. T. Smith,$^{11}$
C. Sorrell$^{9}$
\\
$^{1}$Astronomy Centre, University of Sussex, Falmer, Brighton BN1 9QH, UK\\
$^{2}$Universit\'e de Strasbourg, CNRS, Observatoire astronomique de Strasbourg, UMR 7550, 67000 Strasbourg, France\\
$^{3}$Aix Marseille Univ, CNRS, CNES, LAM, Marseille, France\\
$^{4}$Department of Astronomy, University of Maryland, College Park, MD 20742, USA\\
$^{5}$NASA Jet Propulsion Laboratory, 4800 Oak Grove Dr, Pasadena, CA 91011, USA\\
$^{6}$California Institute of Technology, 1200 E California Blvd, Pasadena, CA 91125, USA\\
$^{7}$NASA Goddard Space Flight Center, Greenbelt, MD 20771, USA\\
$^{8}$Department of Astronomy, University of Massachusetts, Amherst, MA 01003, USA\\
$^{9}$School of Physical Sciences, The Open University, Walton Hall, Milton Keynes, MK7 6AA, UK\\
$^{10}$Max Planck Institut f\"ur Extraterrestrische Physik, Giessenbachstrasse 1, 85748 Garching, Germany\\
$^{11}$Department of Physics \& Astronomy, University of Toledo, Toledo, OH, USA\\
}

\date{Accepted XXX. Received YYY; in original form ZZZ}

\pubyear{2023}

\begin{document}
\label{firstpage}
\pagerange{\pageref{firstpage}--\pageref{lastpage}}
\maketitle

\begin{abstract}
The PRobe far-Infrared Mission for Astrophysics (PRIMA) concept aims to perform mapping with spectral coverage and sensitivities inaccessible to previous FIR space telescopes. PRIMA's imaging instrument, PRIMAger, provides unique hyperspectral imaging simultaneously covering 25--235~$\umu$m. 
%
We  synthesise images  representing a deep, 1500~hr~deg$^{-2}$  PRIMAger survey, with realistic   instrumental and confusion noise.
We demonstrate that we can construct catalogues of galaxies with a high purity ($>95$ per cent) at a source density of 42k~deg$^{-2}$ using PRIMAger data alone.
Using the \texttt{XID+} deblending tool we show that we  measure fluxes with an accuracy better than 20 per cent to flux levels of 0.16, 0.80, 9.7 and 15~mJy at 47.4, 79.7, 172, 235~$\umu$m respectively. These are a factor of $\sim$2 and $\sim$3 fainter than the classical confusion limits for 72--96~$\umu$m and 126--235~$\umu$m, respectively.
At $1.5 \leq z \leq 2$, we detect and accurately  measure fluxes in 8--10 of the 10 channels covering 47--235~$\umu$m for sources with  $2 \la \log({\rm SFR}) \la 2.5$, a 0.5 dex improvement on what might be expected from the classical confusion limit. 
Recognising that  PRIMager  will operate in a context where high quality data will be available at other wavelengths, we investigate the benefits of introducing additional prior information. We show that by introducing even weak prior flux information when employing a higher source density catalogue (more than one source per beam) we can obtain accurate fluxes  an order of magnitude below the classical confusion limit for 96--235~$\umu$m.

\end{abstract}




\section{Introduction}\label{sec:intro}

The process of star formation is integral to understanding galaxy evolution \citep[e.g.][]{Madau2014}. However, a  significant fraction of the UV emission from hot massive stars, which trace star formation, is obscured by dust and re-emitted at far-infrared (FIR) wavelengths \citep{Cardelli1989, Calzetti2000, Burgarella2013}. 

Previous studies have attempted to derive FIR-related properties of galaxies, and correct for dust attenuation, in order to determine physical properties without directly observing in the FIR, e.g. using the IRX-$\beta$ relation to determine IR luminosity \citep{Meurer1999} or using energy-balancing SED fitting procedures to determine star-formation rates (SFR) and dust attenuation \citep{Malek2018}. However, there is no clear agreement within the literature that such approaches are universally applicable or accurate, e.g. with deviations from the IRX-$\beta$ relation being found \citep{Narayanan2018} as well as discrepancies between SFRs and dust attenuation values obtained when fitting SEDs with and without IR photometry \citep{Riccio2021, Pacifici2023}.

Moreover, the total emission we receive from galaxies in the infrared,  the cosmic infrared background \citep[CIB,][]{Puget1996}, forms roughly half of the total extragalactic background light \citep[e.g.][]{Hauser2001, Dole2006}. The discovery of this high CIB, along with wide-field FIR and sub-mm surveys, revealed a population of galaxies which are heavily enshrouded in dust and are known as dusty star-forming galaxies \citep[see][for a review]{Casey2014}. The most luminous DSFGs are thought to be the most intense stellar nurseries in the Universe, with incredibly high SFRs \citep{Rowan2018} and are therefore of crucial importance when it comes to understanding the cosmic star-formation history of the Universe \citep{Long2022}. Observations at short wavelengths can detect DSFGs, but they may be misidentified as unobscured galaxies at higher redshifts \cite[e.g.][]{2023ApJ...943L...9Z}.  Statistical studies of populations of these DSFGs show clearly that their FIR luminosity is significant at most epochs and dominates the luminosity density of the universe at some \cite[e.g.][]{2013MNRAS.432...23G,2013A&A...554A..70B}.

Observations across the FIR wavelength range are therefore required in order to better characterise and constrain the physical properties of galaxies. However, due to the opacity of the atmosphere for much of the FIR wavelength range, these observations must be conducted either from the stratosphere or space. Previous space-based FIR observatories have each significantly enhanced our understanding of the dusty Universe but have also been limited in their capabilities. The first spaced-based telescope to survey the full sky at IR wavelengths was the {\em Infrared Astronomical Satellite} ({\em IRAS}, \cite{Neugebauer1984}), but was only able to conduct shallow observations, detecting only the most luminous IR galaxies (LIRGS). {\em The Infrared Space Observatory}  \cite[{\em ISO}][]{Kessler1996} provided spectroscopy at IR wavelengths \citep[see][for a review]{Genzel2000} but was limited to observing the local Universe due to low sensitivity. Imaging and spectroscopy from the Spitzer Space Telescope \citep{Werner2004} greatly advanced our understanding of obscured star-formation, with the {\em Spitzer/MIPS} 24$\umu$m emission widely used as a tracer of obscured luminosity and SFR \citep{Reddy2008, Elbaz2011, Shivaei2017}. {\em Spitzer}, however, was ultimately limited to only five years of cold mission and its spectroscopy was mainly limited to wavelengths below 35$\umu$m. The {\em Herschel Space Observatory} \citep{Pilbratt2010} significantly extended the survey parameter space, discovering extreme DSFGs 
\cite[e.g.][]{2013Natur.496..329R} and was able to further constrain the CIB \citep{Viero2015, Sides1, Duivenvoorden2020} and the evolution of the IR luminosity function \citep{Gruppioni2010, Gruppioni2013} but could only image in a small number of broad bands. Overall, however, these past missions were limited in imaging to a small number of broad-bands which were not able to capture all the features across the IR range. Previous IR spectroscopy suffered from limited sensitivity or wavelength coverage. 

Imaging data from these previous observatories were also limited in depth due to what is known as confusion noise (particularly {\em Spitzer/MIPS} \citep{Dole2004} and {\em Herschel/SPIRE} \citep{Nguyen2010}). FIR space-based telescopes suffer from poor angular resolution due to the limited mirror sizes, which leads to the blending of sources when the telescope beam is large compared to their average separation. This gives rise to confusion noise, which increases with the observed wavelength for a given mirror size. 

Identifying the need to improve upon our coverage of the FIR sky, NASA released an Announcement of Opportunity (AO) for an Astrophysics Probe Explorer limited to two themes as recommended by the National Academies' 2020 Decadal Review, one of which being a far infrared imaging or spectroscopy mission. In response to this call, The PRobe far-infrared Mission for Astrophysics (PRIMA) concept mission has been developed\footnote{\url{https://prima.ipac.caltech.edu/}}. PRIMA is a 1.8m space-based telescope which will be cryogenically-cooled to 4.5\,K. This FIR observatory has two planned instruments; the Far-Infrared Enhanced Survey Spectrometer (FIRESS) and the PRIMA Imaging Instrument (PRIMAger). FIRESS is a spectrometer covering the 24-235\,$\umu$m wavelength range in 4 grating modules with spectral resolution $R = \lambda / \Delta \lambda \sim 100$. A high resolution mode will allow it to reach $R$ of thousands across full band. The PRIMAger instrument is composed of two bands. The first offers hyperspectral imaging with a $R \sim 10$ providing 12 independent flux measurements from 25 to 80\,$\umu$m while the second provides 4 broad band filters between 96 and 235\,$\umu$m, all sensitive to polarization. Both instruments will operate with 100\,mK cooled kinetic inductance detectors allowing for an incomparable improvement of sensitivity in FIR. As an observatory PRIMA will cover a wide range of science topics such as, but not limited to, origins of planetary atmospheres, evolution of galaxies and build-up of dust and metals through cosmic time \citep{Moullet2023}.

PRIMAger will be able to provide significantly improved sensitivity compared to previous FIR spaced-based imaging instruments, e.g. by over $\sim$2 orders of magnitude for point sources compared to {\em Herschel/PACS} (see Section~\ref{sec:primager-noise} for more details on PRIMAger sensitivity capabilities). However, to realise the full benefit from this sensitivity, it will be essential to reduce the impact of confusion.

Various statistical methods have been developed to overcome the problems presented by confusion when estimating fluxes from FIR maps. One such tool is \texttt{XID+}, developed by \cite{Hurley}. It is a deblending tool which uses a probabilistic Bayesian framework in which to include prior information on galaxy positions and fluxes and to obtain the full posterior probability distribution for fluxes. Positional priors can come from short-wavelength FIR maps, or from catalogues at other (e.g. near-IR) wavelengths. \cite{Hurley} found that \texttt{XID+} performs better on flux accuracy and flux uncertainty accuracy for simulated SPIRE maps than previous prior-based source extraction tools, such as \texttt{DESPHOT} and \texttt{LAMBDAR} (e.g. \texttt{XID+} at 10mJy had similar accuracy to \texttt{LAMBDAR} at 70mJy), and has been utilised in performing source extraction for the Herschel Extragalactic Legacy Project  \cite[HELP;][]{Shirely2019, Shirley2021} and is now a tool used in the wider community  \cite[e.g.][]{2023AJ....165...31S}.

This paper will demonstrate that by utilising the flux modelling capabilities of \texttt{XID+}, accurate flux measurements of galaxies can be obtained below the classical confusion limit from simulated PRIMAger maps. In Section \ref{sec:simulations}, we outline how the simulated PRIMAger maps are generated and the confusion noise estimated. Blind source detection is performed on the maps in Section \ref{sec:catalogues} to produce prior  catalogues with positions to be used to de-blend the confused maps. In Section \ref{sec:xid_and_priors}, we explore how prior information affects the flux modelling of \texttt{XID+}. In Section \ref{sec:results}, we show how \texttt{XID+} performs in terms of measured flux accuracy across the whole simulated dataset and compare to the classical confusion limits. We then discuss the implications of these results on which galaxies we are able to determine the physical properties of in Section \ref{sec:discussions} and make final conclusions in Section \ref{sec:conclusions}.

\section{Simulations}
\label{sec:simulations}

\subsection{Simulated PRIMAger Maps}
\label{sec:sides}

\begin{figure*}
\centering
\begin{tabular}{ccc}
\includegraphics[width=6cm]{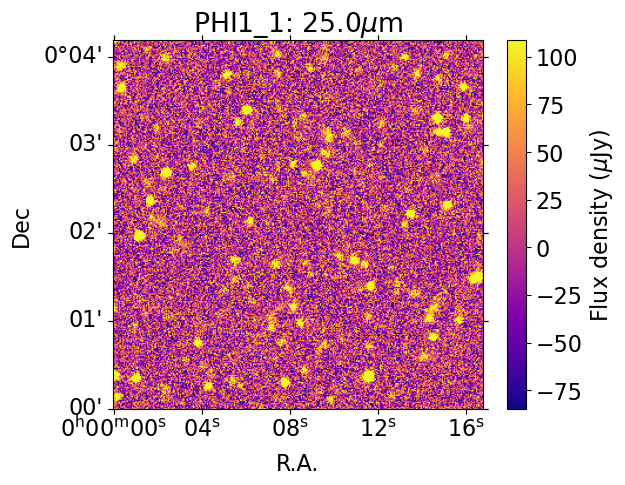} & 
\includegraphics[width=6cm]{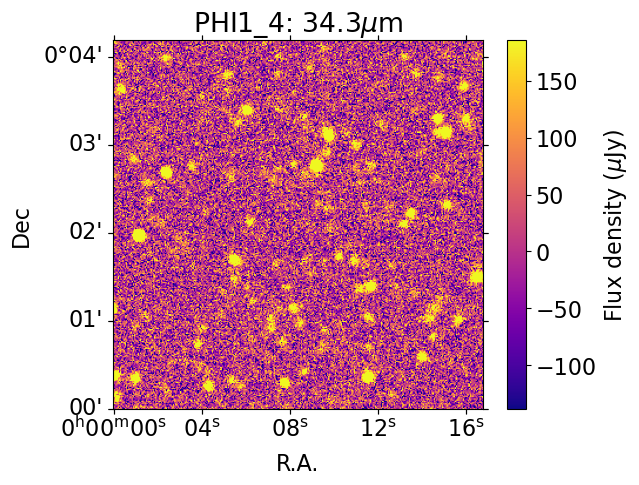} & 
\includegraphics[width=6cm]{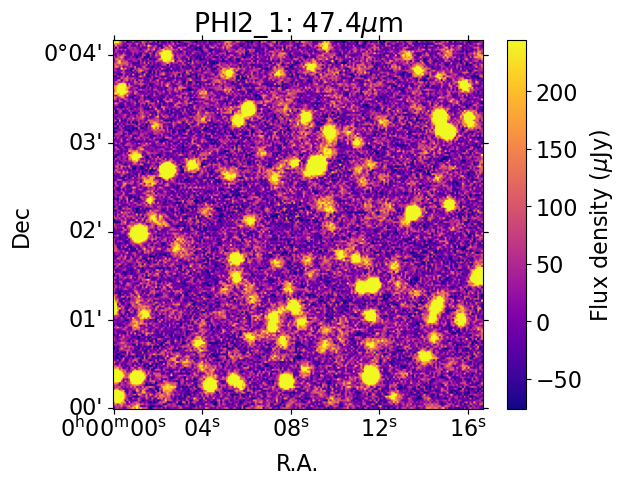}\\
\includegraphics[width=6cm]{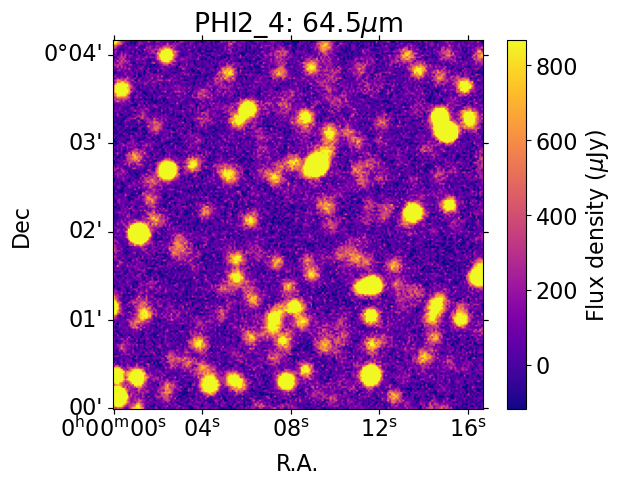} & 
\includegraphics[width=6cm]{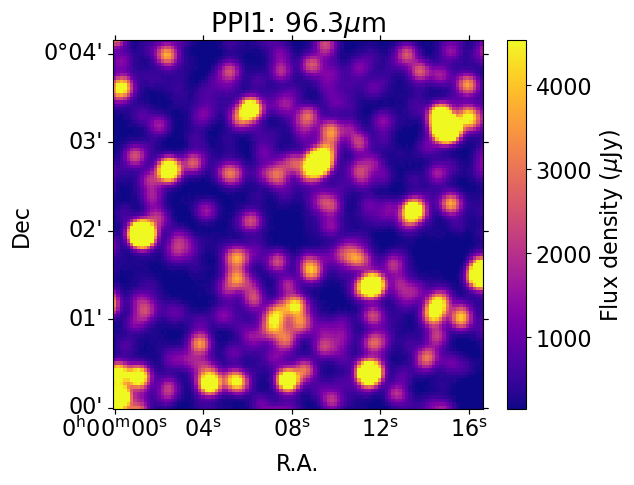} & 
\includegraphics[width=6cm]{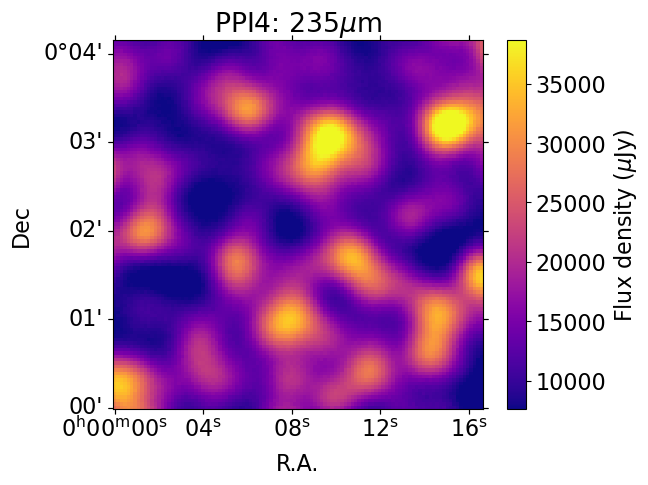}\\
\end{tabular}
\caption{\label{fig:map-cutouts} Simulated PRIMAger maps including both ``instrumental noise'' and source confusion, illustrating the transition from instrumental dominated at short wavelengths to confusion dominated at longer wavelengths. The sources are drawn from the SIDES simulations. The instrumental noise synthesises observations of 1500 hr deg$^{-2}$ and is discussed in Section \protect\ref{sec:primager-noise}. Cutouts are $4\arcmin\times 4 \arcmin$ in representative channels with $R=10$ in bands PHI1 and PHI2 and $R=4$ in band PPI. }

\end{figure*}

To test how well PRIMAger will recover fluxes of sources in the presence of confusion noise, we utilise simulated PRIMAger maps generated by B\'ethermin et al. (2024, submitted) (hereafter referred to as B24) using the Simulated Infrared Dusty Extragalactic Sky (SIDES) simulation \citep{Sides1}. The SIDES simulation, map generation process, estimation of baseline confusion limits and simple blind detection of sources in the absence of instrumental noise are all described and presented in B24, however we summarise some of the relevant information here.

PRIMAger will be able to conduct hyperspectral imaging with linear variable filters in two bands, PHI1 and PHI2, between 25 and 80$~\umu$m with $R \sim 10$. For simplicity, we represent each of these bands with 6 continuous channels spanning the wavelength range of the band (PHI1\_1 to PHI1\_6 for band PHI1 and PHI2\_1 to PHI2\_6 for band PHI2). PRIMAger will also be able to image with polarimetry via 4 broad band channels (PPI1--PPI4), centered at 96, 126, 172 and 235~$\umu$m with $R\sim4$ sensitive to 3 angles of polarisation (see B24 for further discussion of PRIMAger's polarimetry capabilities). Table \ref{tab:filter_stats} includes the central wavelengths and estimated beam full-width half-maxima (FWHM) for each of the 12 representative channels for bands PHI1 and PHI2 as well as the 4 polarimetry channels. PRIMAger will be able to observe simultaneously with all bands, however, due to their relative position on the focal plane, all bands observe different parts of the sky and mapping is needed to cover a region of interest.


SIDES is a simulation of the extragalactic sky in the far-infrared and the millimetre domain, starting from a dark-matter halo light cone with galaxy properties generated using a semi-empirical model. It is able to reproduce a large set of observed galaxy properties such as the source number counts at various angular resolutions, the redshift distributions and the large-scale anisotropies of the CIB. For this work, the latest version of the simulation presented in \cite{Sides2} is used. 



The output from the SIDES simulation is a lightcone catalogue of 1.4~deg $\times$ 1.4~deg, $0< z <10$, corresponding to a comoving volume of 0.17~Gpc$^{3}$ containing 5.6M galaxies. The catalogue also contains the fluxes of each source which are obtained by integrating the spectral energy distribution of the SIDES galaxies over the representative PRIMAger channels. Simulated maps which contain confusion noise but no instrumental noise are generated for each of the 16 channels (hereafter these maps are referred to as `noiseless maps', with simulated instrumental noise being added to produce `noisy maps' as described in Section \ref{sec:primager-noise}). The noiseless maps are generated by attributing the flux of the sources to the centre of the pixels at which they are located and then convolving the map with the relevant beam profile. The beam profiles are assumed to be Gaussian with FWHM values given in Table~\ref{tab:filter_stats}. Map pixel sizes are 0.8, 1.3 and 2.3 arcsec for bands PHI1, PHI2 and PPI, respectively. Cutouts of the same region from the simulated PRIMAger maps in 6 of the channels are shown in Figure~\ref{fig:map-cutouts}, with the effect of confusion noise clearly demonstrated as you move to longer wavelengths, whereby sources become increasingly blended. The estimation of the confusion noise from the maps containing no instrumental noise is discussed below.

\subsubsection{Estimating Classical Confusion Noise}
\label{sec:primager-confusion}

Confusion noise arises due to surface brightness fluctuations in the maps arising from the astronomical sources themselves, convolved with the telescope beam.  The lowest flux at which an individual point-like source can be identified above that fluctuating background is called the confusion limit \citep{Condon}. In order to estimate the confusion limit of a simulated noiseless PRIMAger map, B24 applied a $5\sigma$-clipping process. 
The standard deviation, $\sigma$, of all pixels is computed, 5$\sigma$ positive outliers are masked and the standard deviation of the unmasked pixels is recomputed. This process is iterated until the standard deviation converges, giving the $1\sigma_{\rm conf}$ confusion noise. The {\em classical} confusion limit is then defined as 5 times this confusion noise, values of which are given in Table~\ref{tab:filter_stats} for each of the PRIMAger channels.


\subsection{PRIMAger Sensitivities and Simulated Noise}
\label{sec:primager-noise}

With significantly improved sensitivity compared to previous FIR space-based telescopes, PRIMAger will reach a point source sensitivity of 220 and 300~$\umu$Jy in bands PHI1 and PHI2, respectively, at the 5$\sigma$ level with integration time 10~hr~deg$^{-2}$. Likewise, it will reach point source sensitivities of 200, 300, 400 and 500~$\umu$Jy in bands PPI1, PPI2, PPI3 and PPI4, respectively.  For comparison, the {\em deepest} surveys with {\em Herschel} at 110, 160 and 250~$\umu$m (close to the PPI2, PPI3, and PPI4 bands)  reached 5$\sigma$ sensitivities of 1100, 2100, 3800 $\umu$Jy in \cite[see table 5 in][] {2012MNRAS.424.1614O}

In order to make realistic PRIMAger maps, we add instrumental\footnote{The non-confusion noise in PRIMager maps arises from multiple sources, including the detectors and photon statistics, for convenience we aggregate these and refer to them as ``instrumental noise''.} noise to the simulated maps.  We assume a deep, 1500 hr~deg$^{-2}$ survey which is expected to give 5$\sigma$ point source sensitivities of 88, 108, 29, 45, 67 and 82~$\umu$Jy in bands PHI1, PHI2, PPI1, PPI2, PPI3 and PPI4 respectively. 

We add Gaussian noise to each pixel based on the nominal point source sensitivities (5$\sigma_{\text{inst}}$) for the considered survey design for each PRIMAger channel in Table~\ref{tab:filter_stats}. We assume no spatially correlated noise and that the instrumental noise is constant across each respective map. Maps which contain both confusion noise and this added instrumental noise are referred to as `noisy maps' for the remainder of the paper. 

\begin{table*}
\raggedright
\caption{Properties and detection limits in the 12 representative PRIMAger channels for the two hyperspectral bands, PHI1 and PHI2, and the four polarimetry broad band channels, PPI1-PPI4. Beam FWHMs are estimated (column 3) for the  baseline telescope aperture (1.8 m) and detector and pixel layout. The point source sensitivities are given (column 4) for a deep survey  observed for 
$\sim$1500~hr~deg$^{-2}$ in the absence of confusion. The classical confusion limit as estimated by B24 is also quoted (column 5) for each channel. This is defined as 5 times the confusion noise which is obtained by estimating the variance in each of the maps via an iterative clipping process in the absence of instrumental noise.  The depth of each Wiener-filtered catalogue of monochromatic, blind detections at the 95 per cent purity level are presented (column 6) with  details  discussed in Section~\protect\ref{sec:wiener_cat}. The depth reached by the two runs of \texttt{XID+} with two different prior catalogues are also provided (columns 7 and 8) and are discussed in Section \protect\ref{sec:results}. The {\em Wiener-filtered} prior catalogue is self-consistently derived from Wiener-filtered catalogues extracted from the synthetic data, the {\em Deep} prior catalogue comes from the input model and represents a prior catalogue from other observatories with weak flux priors. \texttt{XID+} depths are the limiting fluxes defined in equation \protect\ref{eq:limiting_flux}.  N.B. the flux accuracy tolerance in the purity analysis of the Wiener-filtered catalogue is different from than used in the definition of XID+ limiting flux. Thus, their values are not directly comparable. Data are quoted to 3 significant figures. }
\label{tab:filter_stats}
\begin{tabular}
{|p{1.0cm}|p{1.4cm}|p{1.4cm}|p{2.0cm}|p{2.0cm}|p{2.0cm}|p{2.0cm}|p{2.0cm}|} 
\hline

Channel & Central Wavelength & Estimated Beam FWHM &  Sensitivity (5$\sigma_{\text{inst}}$) & Classical Confusion (5$\sigma_{\text{conf}}$) & Wiener-filtered Catalogue Depth (95\% purity) & XID+ Depth, Wiener-filtered Prior ($5\sigma_{\rm MAD}$) & XID+ Depth, Deep prior, ($5\sigma_{\rm MAD}$)\\
& [$\umu$m] & [\arcsec] & [$\umu$Jy] & [$\umu$Jy] & [$\umu$Jy] & [$\umu$Jy] & [$\umu$Jy] \\
\hline
PHI1\_1    &25.0       &4.1      & 70 &20   & 55 & 93 & 82\\
PHI1\_2    &27.8       &4.3      & 79 & 27  & 61 & 107 & 86\\
PHI1\_3    &30.9       &4.6      & 88 & 37  & 69 & 121 & 93\\
PHI1\_4    &34.3       &4.9      & 99 & 51  & 79 & 138 & 108\\
PHI1\_5    &38.1       &5.2      & 114 & 71 & 94 & 162 & 116\\
PHI1\_6    &42.6       &5.7      &134 & 107 & 117 & 181 & 149\\
PHI2\_1    &47.4       &6.2      &83 & 161  & 87  & 163 & 95\\
PHI2\_2    &52.3       &6.7      & 94& 249  & 115 & 209 & 117\\
PHI2\_3    &58.1       &7.3      &108 & 401 & 160 & 271 & 138\\
PHI2\_4    &64.5       &8.0      &123 &667  & 225 & 387 & 167\\
PHI2\_5    &71.7       &8.8      &153 &1120 & 336 & 602 & 229\\
PHI2\_6    &79.7       &9.7      &172 &1850 & 521 & 801 & 285\\
PPI1    &96.3       &11.6      &29 & 4520   & 770 & 1680 & 281\\
PPI2    &126       &15.0      &45 &12300    & 2002 & 4090 & 747\\
PPI3    &172       &20.3      &67& 28400    & 5037 & 9700 & 2650\\
PPI4    &235       &27.6      &82& 46000    & 18023 & 14600 & 7030\\

\hline
\end{tabular}
\end{table*}


\section{Source Detection}
\label{sec:catalogues}

\texttt{XID+} deblends maps using the positions of known sources (Section \ref{sec:xid}) and therefore requires a catalogue containing their prior positions. Usually, such a prior catalogue is obtained from shorter wavelength ancillary data from other telescopes, as was done by \cite{Pearson2018} who used optical data in the COSMOS field to deblend {\em Herschel/SPIRE} maps with \texttt{XID+}. However, to demonstrate that it is possible to detect sources and accurately measure their fluxes entirely from PRIMAger data in a self-contained way, a source detection process is run on the simulated PRIMAger maps themselves. This is possible due to the wide spectral coverage of PRIMAger, particularly in the PHI1 band (25--43~$\umu$m). This band is not limited by confusion and allows for sources to be detected at multiple wavelengths. Additionally, this band will capture PAH emission lines from star-forming sources around cosmic noon, enhancing their detection probability. 

In order to explore the impact of different prior knowledge on the flux accuracy  of \texttt{XID+} (Section \ref{sec:prior_knowledge}), we consider two different source detection methods.

\subsection{Blind Detection Without Instrumental Noise}
\label{sec:blind_cat}

A blind source detection algorithm performed on all of the noiseless PRIMAger maps by B24, who produced a catalogue of 101,540 sources from the 1.96~deg$^{2}$ maps. The basic algorithm they employ searches for local maxima within a $5\times 5$ pixel region. The threshold was set to be 5 times the measured confusion noise.

\subsection{Blind Detection On Wiener Filtered Maps With Instrumental Noise}
\label{sec:wiener_cat}

In far-IR and sub-millimeter blind surveys, it is common to perform blind detection by cross-correlating the signal with the PSF of the instrument. This method is expected to maximize the signal-to-noise ratio (S/N) on isolated point sources with white noise. This is appropriate for shallow surveys, dominated by instrumental noise. However, for the deep observations planned for PRIMA on extra-galactic deep fields, the spatially correlated confusion noise is non-negligible even in the shortest PHI1 bands and starts to dominate the noise in the data at PHI2 and longer wavelengths. PSF matching filter, in this case, no longer maximises the S/N of point sources, rather it increases the confusion noise and reduces the completeness of blind detection.

To maximise the S/N of blind source detection in data with substantial confusion noise, previous blind far-IR and millimetre surveys have introduced a Wiener filter as the general form of matching filter kernel that optimise point source blind detection on confusion-limited data \citep{Chapin2011, Geach2017, Shirley2021}. The philosophy of this method is a compromise between the uncorrelated white instrumental noise (which benefits from a wider kernel) and other spatially correlated confusion noise (which benefits from a narrower kernel and local background removal) to maximise the signal-to-noise ratio of point source blind detection. Our construction of Wiener filter follows the principles in \citet{Chapin2011} and the similar frameworks of constructing blind catalog in HELP project \citep{Shirley2021}. We refer the readers to those references for details but summarise the outcomes of the match-filtering as follows. 

\begin{figure}
    \centering
    \includegraphics[width=\linewidth]{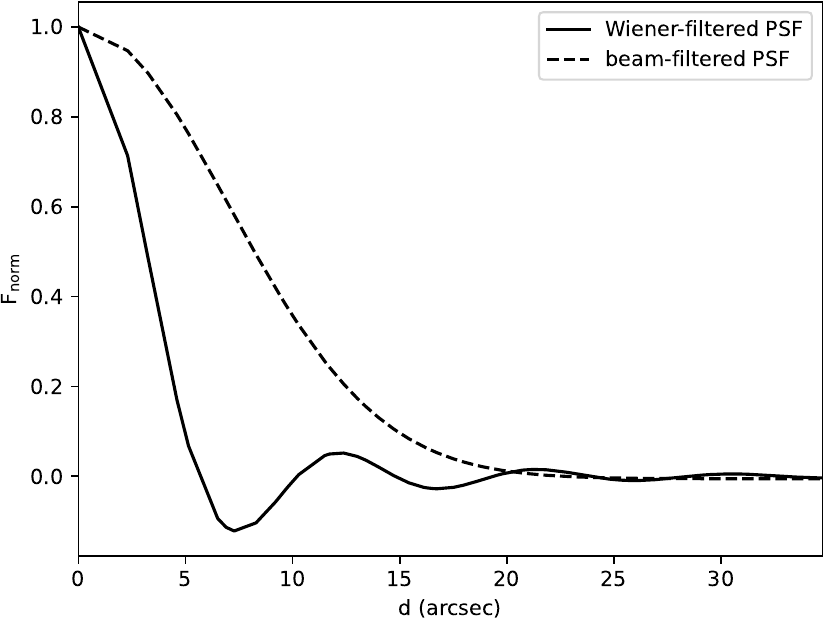}
    \caption{The effective Point Spread Function (PSF) profile in a PPI1 image after applying: a Wiener filter (solid line); or a n\"aive PSF filter (dashed line). The n\"aive filter is optimal for flux estimation of isolated point sources, including non-local data with reduced weight to enhance the signal to instrumental noise ratio in comparison to a central, purely local, estimate.  The Wiener filter has a much narrow central beam which provides some reduction in instrumental noise but balanced against adding in confusion noise from non-local data.  The Wiener filter also has a negative feature (at around 7\arcsec\ in this example) which provides a local background subtraction, actively reducing confusion noise.}    
    \label{fig:psf_profile}
\end{figure}

We consider the total noise in the simulated data from PRIMAger observations as two main components: a white noise component coming from the instrumental noise and a confusion noise component coming from other point sources in the map. In each band, we take the instrumental and confusion noise level expected for 1500-hour PRIMAger deep survey over 1 deg$^2$, create the Wiener filters following \citet{Chapin2011} and derive the corresponding match-filtered map. A comparison between the effective PSF profile after applying Wiener filter and instrument PSF filter to the simulated PRIMA observation is illustrated in Fig.~\ref{fig:psf_profile}. The effective PSFs after Wiener filtering have primary peaks narrower than the instrument PSF, this reduces source blending and improve the completeness of blind source detection in confusion-dominated PHI2 and PPI1-PPI4 bands. The higher order ringings feature from Wiener filtering introduces additional fake sources around in blind detection, which we identify and remove later. However, the impact of ringing are limited to regions around very bright sources and the corresponding fake sources could be removed based our knowledge on their relative intensity compared to the nearest bright sources. 

The blind source detection in the match-filtered maps is made using the find\_peak method provided by \texttt{photutils}. A source is identified if the central pixel is the maximum among all pixels in a $5\times5$ pixel region. The maps are calibrated following \citet{Chapin2011} in mJy/Beam such that the point source flux could be estimated directly from the peak. 

To remove false sources created by Wiener filtering we examined all sources in the simulation that are bright enough to produce ringing features above the total noise level of the map, 
$\sigma_\mathrm{total}$, ($\sigma_\mathrm{total}^2 = {\sigma_\mathrm{inst}^2 + \sigma_\mathrm{conf}^2}$, where $\sigma_\mathrm{inst}$ and $\sigma_\mathrm{conf}$ are the instrumental noise and the confusion noise, respectively).
We then predict the expected intensity of the corresponding ringing features. Sources with fluxes less than five times the expected ringing feature intensity are considered as contaminated and removed from the blind detection catalog. We note that although this conservative cut could also remove some real faint sources this will be reflected in our completeness and flux accuracy estimates and further optimisation could improve our results.

Before constructing the prior list, it is critical to define a cut on the depth for blindly detected source catalogs to avoid significant contamination from false detections, while maintaining high completeness. Significant contamination could be a problem not because of the false objects themselves but also, through the \texttt{XID+} modelling,  reduce the flux accuracy for real sources. 

 Far-IR and submillimeter surveys usually set a flux cut based on the purity derived from statistical analysis. Purity is defined as the fraction of detected source above certain flux limit that have corresponding  counterparts in the simulated input catalogue close enough in positions and fluxes. In our analysis, we consider that a correct counterpart to a blindly detected source satisfies the following criteria (similar to B24): 
 \begin{enumerate}
 \item the positional offset between blindly detected source and the counterpart,  $d_\mathrm{off}$, satisfies $d_\mathrm{off} \le \theta_\mathrm{FWHM}/2$, where $\theta_\mathrm{FWHM}$ is the FWHM of the instrument PSF
 \item the observed flux (S$_\mathrm{obs}$) of the blindly detected source and the true flux (S$_\mathrm{True}$) of the counterpart satisfies      $S_\mathrm{True}/2 \le S_\mathrm{obs} \le 2 S_\mathrm{True}$
\item the counterpart is the brightest source that satisfies criteria (i) and (ii).
 \end{enumerate}

We crossmatch the blind detected source catalog with the simulation input catalogues using those criteria. For each band, we perform crossmatching on blindly detected sources with S/N$>$2.5. The purity of blindly detected sources above different flux limits are then derived accordingly. We choose a cut in observed flux corresponding to the 95 per cent purity  and the flux threshold for each band are listed in Table.~\ref{tab:filter_stats}. The resulting blindly detected single-band catalog reaches completeness of $\sim$ 83 per cent at PHI1 bands, $\sim$ 67 per cent at PHI2 bands and $\sim$ 75 per cent at PPI1-PPI4 bands on sources with S$_\mathrm{True} > \sigma_\mathrm{total}$. These catalogs are further cross-matched from the shortest to the longest wavelength to obtain a unique list of priors from Wiener filtering.
\section{XID+ and Prior Information}
\label{sec:xid_and_priors}

\subsection{XID+: A Probabilistic De-Blender}
\label{sec:xid}

\texttt{XID+}\footnote{\url{https://github.com/H-E-L-P/XID_plus}}, developed by \cite{Hurley}, is a prior-based source photometry tool which is able to simultaniously estimate the fluxes of a collection of sources with known positions. The basic model of \texttt{XID+} assumes that the input data ($\boldsymbol{d}$) are maps with $n_{1} \times n_{2} = M$ pixels, where the maps are formed from $N$ known sources, with flux densities $\boldsymbol{S_{i}}$ and a background term accounting for unknown sources. The point response function (PRF, $\boldsymbol{P}$) quantifies the contribution each source makes to each pixel in the map and is assumed to be a Gaussian. The map can therefore be described as follows:
\begin{equation}
\boldsymbol{d} = \sum_{i=1}^{N} \boldsymbol{P}\boldsymbol{S_{i}} + N(0, \Sigma_{\text{inst}}) + N(B, \Sigma_{\text{conf}}),
\label{eq:xid}
\end{equation}
where the two independent noise terms represent the instrumental noise and the residual confusion noise which is modelled as Gaussian fluctuations about $B$, a global background. \texttt{XID+} undertakes an MCMC sampling from this probabilistic model to obtain the full posterior. Originally, \cite{Hurley} utilised the Bayesian inference tool, \texttt{Stan}, to perform the MCMC sampling. However, here we implement the \texttt{Numpyro} backend which is built into \texttt{XID+} as it is faster.

The original \texttt{XID+} applied a flat, uniform prior on the source fluxes (from zero flux to the highest pixel value in the map). However, later works \citep{Pearson2017, Pearson2018, Wang2021} demonstrated that by applying more informative flux priors, e.g. from SED-fitting of ancillary photometry, provided improvements in flux accuracy and allowed fainter fluxes to be reliably measured. 

We would expect the choice of prior information provided to affect the modelling accuracy. In the basic \texttt{XID+} model described above, the possible prior information to include are (a) the positions of previously detected sources (i.e. the density of sources) and (b) the prior probability distributions of their  fluxes. The following section investigates the impact of varying these two prior information dimensions on the flux modelling accuracy of \texttt{XID+}.

\subsection{Impact of Prior Knowledge}
\label{sec:prior_knowledge}

In order to investigate the impact of the inclusion of prior knowledge on the modelling accuracy of \texttt{XID+}, we consider (a) varying the density of sources included in the prior source position catalogue as well as (b) varying the prior flux distribution. One would expect the flux modelling accuracy to improve as the density of the prior source position catalogue increases, as the more faint sources are included in the modelling, the fewer sources remain to contribute confusion. However, without any prior flux knowledge, there would be an upper limit, and even a reversal, to the gain in modelling accuracy as the number of prior sources increases due to degeneracies being introduced to the model. With prior flux knowledge, these degeneracies may be overcome. 

We consider three prior source position catalogues of varying source densities:
\begin{enumerate} 
    \item \textbf{B24 Catalogue}: Discussed in Section \ref{sec:blind_cat}
    \item \textbf{Wiener-filtered Catalogue}: Discussed in Section \ref{sec:wiener_cat}
    \item  \textbf{Deep Catalogue}: Here we apply a simple flux cut to the full SIDES simulation catalogue, keeping sources which have a flux greater than 1~$\umu$Jy in the PHI1\_1 band. This produces a catalogue of 588,550 sources, corresponding to a source density of $\sim3.5$ sources/beam in the PPI1 channel.
\end{enumerate}
A summary of these catalogues is presented in Table \ref{tab:cat_stats}. Note that when considering application to real data (i) could, in principle, be generated from PRIMager map data if sufficiently deep that instrumental noise was negligible (ii) could be generated from the PRIMager survey data we are considering here (iii) would require catalogues generated from data from other telescopes.

\begin{table}
\raggedright
\caption{Number of sources  in each of the three prior source catalogues over 1.96~deg$^{2}$  explored in Section \ref{sec:prior_knowledge}. B24 uses a basic peak detection on noise-free (confusion only) maps; the Wener-filtered catalogue is an extraction from a simulation of the deep, 1500~hr~deg$^{-2}$, survey; the Deep catalogue comes from the simulated input catalogue and represents a prior catalogue from other facilities. The source density is given as number of sources per band PPI1 beam. The corresponding flux depth from the simulation input catalogue at this source density is also provided.}
\label{tab:cat_stats}
\begin{tabular}{|p{1.7cm}|p{1.7cm}|p{1.7cm}|p{1.7cm}|} 
\hline
Name & No. of Galaxies & Source Density & Flux Depth\\
& & [sources/beam] (in band PPI1) & [$\umu$Jy]\\
\hline
B24    &101,540       &0.60      &270\\
Wiener-filtered    &82,575       &0.49      &369\\
Deep    &588,550       &3.50     &12.9\\

\hline
\end{tabular}
\end{table}

\begin{figure}
    \centering
    \includegraphics[width=\linewidth]{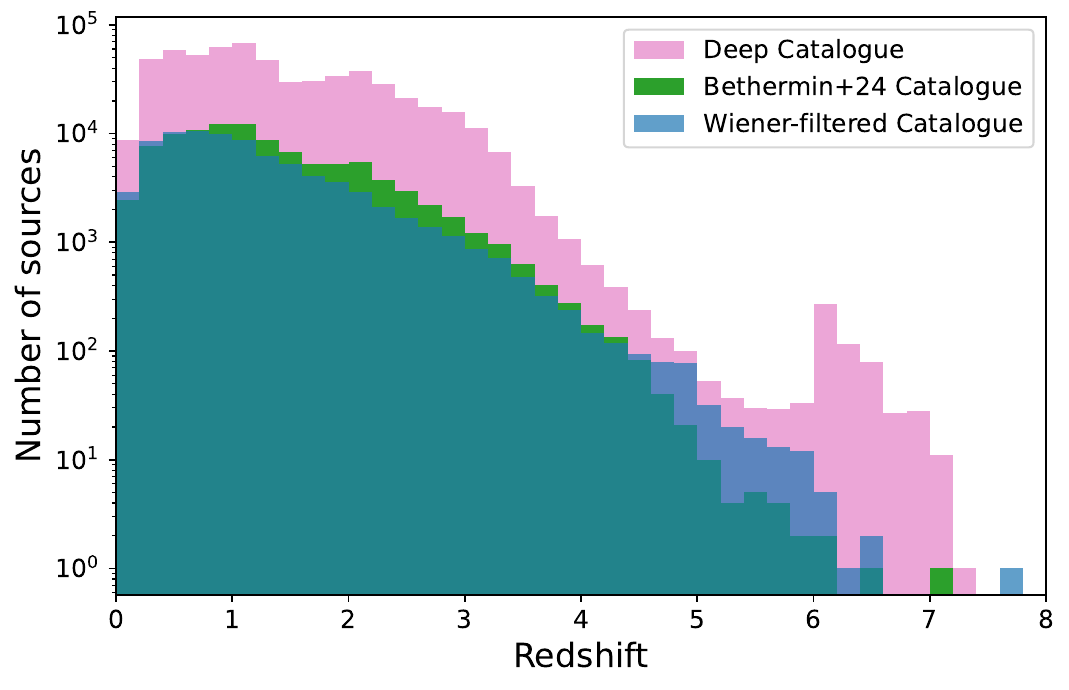}
    \caption{Redshift distribution for the three prior source catalogues considered in Section \ref{sec:prior_knowledge}: The Deep catalogue (pink) of $\sim$590,000 sources which all have  a flux
greater than 1$\umu$Jy in the PHI1\_1 channel, the catalogue of blindly detected sources in the noiseless maps (green) from B24 of $\sim$102,000 sources and the catalogue of blindly detected sources in the Wiener-filtered, noisy maps (blue) with $\sim$83,000 sources.}    
    \label{fig:z_cats}
\end{figure}

\begin{figure*}
    \centering
    \includegraphics[width=\linewidth]{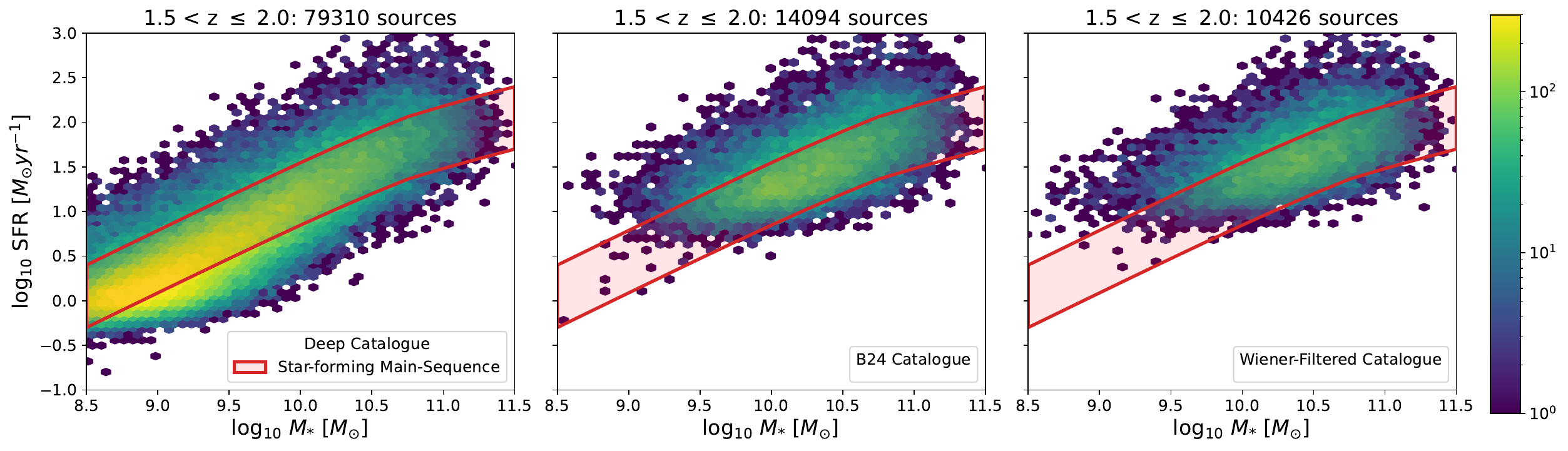}
    \caption{Location of sources within the three catalogues considered in Section \ref{sec:prior_knowledge} on the stellar mass-SFR plane for a single redshift bin of $1.5 < z \leq 2.0$. For reference we indicate a range of reported locations for the star-forming ``main sequence'' in the literature \citep{Speagle2014, Pearson2018, Leslie2020, Leja2022}. The number of sources in each M$_{*}$--SFR bin is shown by the colour scale for the Deep, B24 and Wiener-filtered catalogues in the left, middle and right panels, respectively.}    
    \label{fig:MS_cats}
\end{figure*}

The redshift distributions of the sources in each of the above catalogues is shown in Figure \ref{fig:z_cats}. A secondary peak of sources in the deep catalogue is present at $z \gtrsim 6$ due to the 3.3$\umu$m PAH emission line moving into the PHI1\_1 channel which is used for the selection of sources for this particular catalogue. Figure \ref{fig:MS_cats} shows where the sources from each catalogue lie in the SFR-stellar mass plane for a single redshift bin ($1.5 < z \leq 2.0$) compared to values for the star-forming main-sequence (MS) from the literature \citep{Speagle2014, Pearson2018, Leslie2020, Leja2022}. The Deep catalogue contains significantly more low-mass galaxies as well as a larger population of  galaxies just below the MS, moving towards the quiescent region, across all masses. Conversely, the Wiener-filtered and B24 catalogues have a higher percentage of their total sources above the MS. 

\begin{figure}
    \centering
    \includegraphics[width=\linewidth]{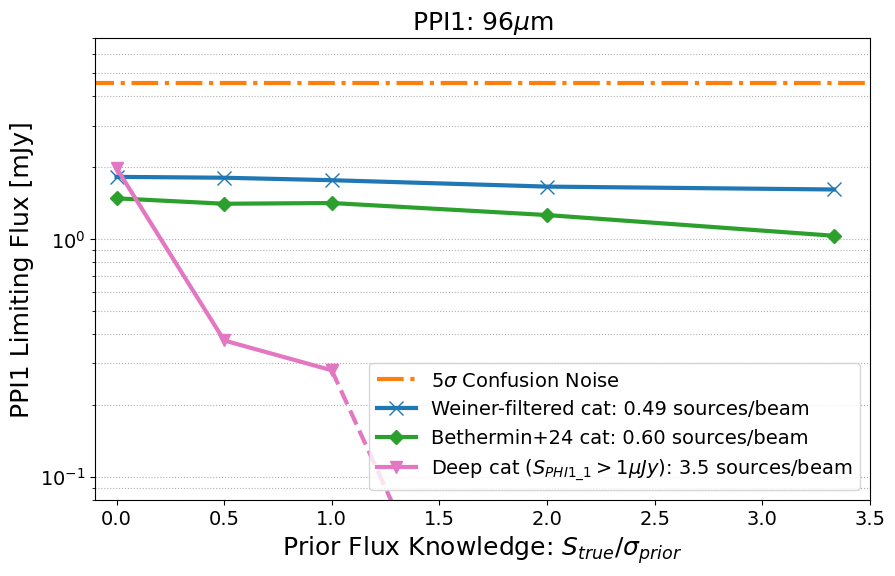}
    \caption{The limiting flux, as defined in Section \ref{sec:limiting_flux}, reached by \texttt{XID+} in the PPI1 channel as a function of the prior flux knowledge. Prior knowledge is defined as the true flux of the sources, $S_{\text{true}}$ divided by the spread on the Gaussian flux prior, $\sigma_{\text{prior}}$ (i.e. as the prior flux knowledge increases, the spread on the flux prior decreases). Results are shown for a sample of the data (totalling $\sim$0.12~deg$^{2}$) from the three prior source catalogues described in Section \ref{sec:prior_knowledge}: Weiner-filtered catalogue (green line with cross markers); blind detected catalogue from B24 (blue line with diamond markers); and  the Deep catalogue (pink line with triangle markers). For the Deep catalogue beyond
    $ S_{\text{true}}/\sigma_{\text{prior}} > 1$ (not plotted, but indicated by the dashed line)
    the limiting flux is $\sim 1\umu$Jy, i.e. flux of the faintest source, indicating that the modelling is performing as well as possible.  Source densities for each of the catalogues are indicated in the legend. The orange dash-dotted line shows the classical confusion limit for the PPI1 channel estimated by B24.}    
    \label{fig:prior_knowledge}
\end{figure} 

For each of the catalogues, \texttt{XID+} is run on a sample of the data covering $\sim 0.12$~deg$^{2}$ with uninformative, flat flux priors (i.e. with uniform flux priors on all sources ranging from zero to the highest pixel value in the respective map.) for the PPI1 channel. This channel is chosen as it is confusion-dominated (i.e. the instrumental noise is negligible compared to the confusion noise) but remains key for many of the PRIMA science goals. Additionally, \texttt{XID+} was run with Gaussian flux priors centred on the sources' true flux,  with standard deviations of 2.0, 1.0, 0.5, 0.3 times the true flux of the sources (i.e. with increasing prior flux knowledge). In a real survey, these flux prior constraints would likely come from predicted fluxes obtained from SED-fitting procedures utilising ancillary photometry (Section \ref{sec:prior_flux_info_disucssion} discusses this further). The method used for measuring the performance of the flux modelling from \texttt{XID+} for the above runs as well as the subsequent results are described in the following section.

\subsubsection{Limiting Flux Statistic}
\label{sec:limiting_flux}

In order to quantify the flux accuracy of \texttt{XID+} for the varying prior knowledge parameters, we define the following statistics to describe the `limiting flux' reached in each of the PRIMAger maps. Firstly, we quantify the deviation of the extracted fluxes, $S_{\text{obs}}$ from the true fluxes, $S_{\text{true}}$, within bins of true flux using the median absolute deviation (scaled to a Gaussian), $\sigma_{\text{MAD}}$:
\begin{equation}
\sigma_{\text{MAD}}(S_{\text{true}}) = 1.4862 \cdot \text{Median}\left(\frac{\Delta S}{S_{\text{true}}} - \text{Median}\left(\frac{\Delta S}{S_{\text{true}}}\right)\right),
\label{eq:mad}
\end{equation}
where $\Delta S = S_{\text{obs}} - S_{\text{true}}$. We then define the limiting flux, $S_{\text{limiting}}$, as the flux at which $\sigma_{\text{MAD}}$ equals 0.2:
\begin{equation}
S_{\text{limiting}} = S_{\text{true}}\bigg|_{\sigma_{\text{MAD}} = 0.2}.
\label{eq:limiting_flux}
\end{equation}  
This corresponds to the true flux at which the median deviation of the observed fluxes from the true values equals 20\% of the true flux. The choice of this statistic and whether it is a reasonable measure of the flux down to which source fluxes can be accurately recovered is considered in Appendix \ref{sec:choice_of_statistic}.

The prior flux knowledge is quantified as the true flux, $S_{\text{true}}$, over the dispersion in the Gaussian flux prior, $\sigma_{\text{prior}}$. It is worth noting that the $S_{\text{true}}/\sigma_{\text{prior}}\sim3.3$ flux prior is only considered in order to investigate the upper bound of this parameter space, as if we were able to constrain the flux a priori this accurately then new data would add little! 

Figure \ref{fig:prior_knowledge} shows the limiting flux in the PPI1 channel as a function of the prior flux knowledge for all three prior catalogues. It highlights how increasing the prior flux knowledge for the shallower prior source catalogues (the B24 and the Wiener-filtered catalogues) provides negligible gains.
Therefore, it is better to only apply a flat flux prior for prior source catalogues of source densities $< 1$. Increasing the prior flux knowledge at these source densities returns little gain but will likely introduce more assumptions into the modelling, depending on how the prior flux information is obtained. For deeper, higher source density prior catalogues, however, even weak information from flux priors can lead to substantial gains in the \texttt{XID+} flux modelling accuracy and limiting flux. 

\subsection{Choice of Prior Source Catalogue}

To investigate the flux modelling performance of \texttt{XID+} across the full simulated PRIMAger dataset, we will continue with both the Wiener-filtered and the Deep prior source catalogues. The former will be used as the benchmark as it is generated from the more realistic maps which include instrumental noise, providing a  robust and conservative estimate of a realistic blind source detection process. No prior flux information will be used with this catalogue as to avoid introducing assumptions for little gain. This run will provide the most conservative limiting flux results. 

The Deep prior source catalogue is not generated from PRIMAger's capabilities or from the maps themselves. However, a catalogue of such source density and depth is possible to obtain from wide-field surveys conducted by higher resolution observatories, such as The Nancy Grace Roman Space Telescope. It is important to understand how much can be gained from utilising such rich ancillary datasets. Additionally for this run, we include prior flux information for each source as a Gaussian distribution centered on the true flux, $S_{\text{true}}$, with a spread of $\sigma_{\text{prior}} = S_{\text{true}}$. This is to test the flux modelling performance in the more informative prior knowledge regime.

The blind detection catalogue produced by B24 represents what is possible to achieve in the limit of no instrumental noise (i.e. for very deep surveys). However, due to it being produced from the noiseless simulated PRIMAger maps, rather than the more realistic maps with added simulated noise which \texttt{XID+} will be run on, it will not be considered further.

\section{Results}
\label{sec:results}

\subsection{XID+ Photometry with Wiener-filtered Prior Catalogue}
\label{sec:xid_results}

\begin{figure}
    \centering
    \includegraphics[width=\linewidth]{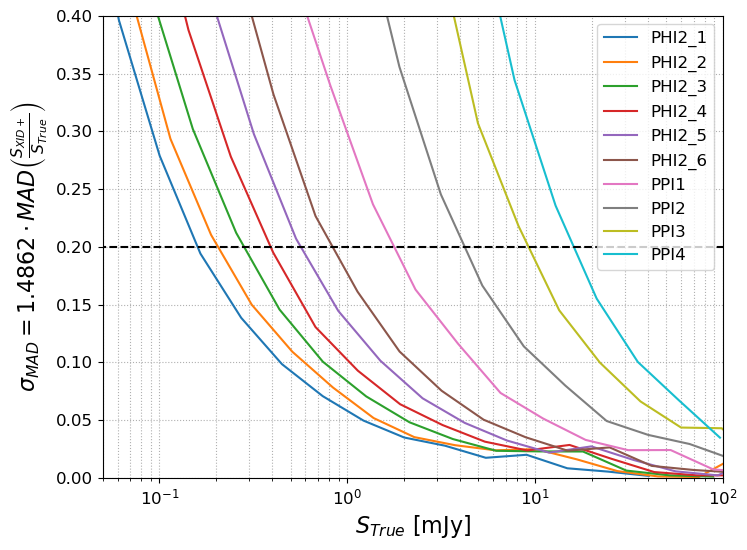}
    \caption{\texttt{XID+} flux accuracy as a function of true flux for Wiener-filtered prior in the 10 reddest PRIMAger channels (coloured solid lines). Flux accuracy is quantified as the scaled Median Absolute Deviation (MAD), $\sigma_{\text{MAD}}$ as defined in equation \ref{eq:mad}, of the ratios of measured source fluxes from \texttt{XID+}. The horizontal black dashed line shows the `limiting flux' threshold at $\sigma_{\text{MAD}} = 0.2$, which represents measured flux accuracy of 20 per cent ($5\sigma$). The true flux at which the coloured solid lines intercept with this threshold is taken to be the 'limiting flux' for the given channel, as defined in equation  \ref{eq:limiting_flux}.}
    \label{fig:mad_vs_true}
\end{figure}

\begin{figure*}
    \centering
    \includegraphics[width=\linewidth]{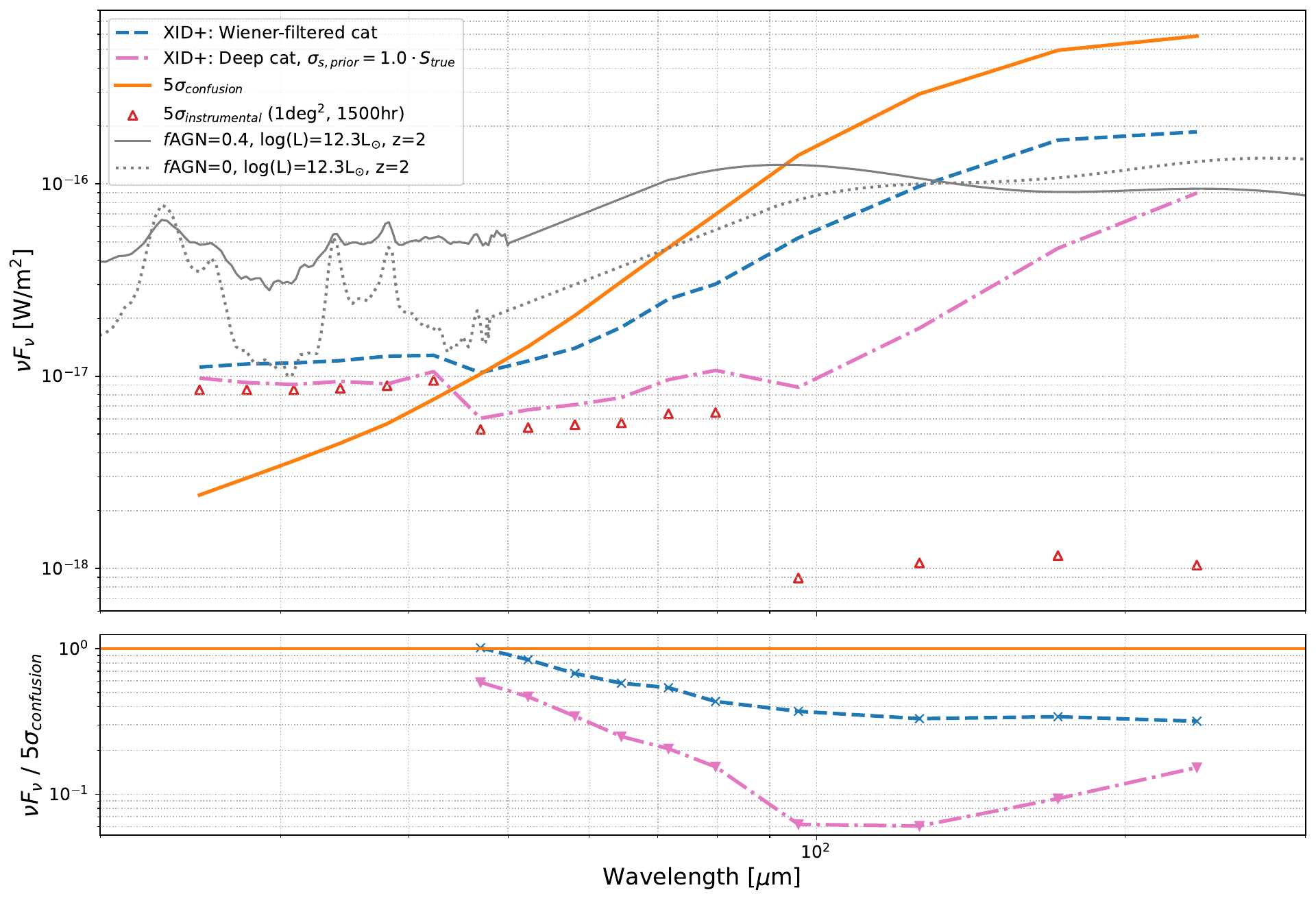}
    \caption{Limiting flux density as a function of wavelength from 25--235$\umu$m for \texttt{XID+} deblending. De-blending with positional and weak flux priors consistently attains $5\sigma$ depths more than an order of magnitude fainter than the classical confusion limit at $\lambda>100$\,$\mu$m. This figure shows that using \texttt{XID+}, SEDs from typical galaxies at $z=2$ can be measured to $\lambda=126$\,$\mu$m using only positional priors (derived from the Wiener-filtered map), and out to the longest PRIMAger PPI channel ($\lambda=235$\,$\mu$m) with the addition of a weak intensity prior. \textbf{Top:} 
    Limiting flux density as a function of wavelength covering the 12 representative channels of the two LVF PRIMAger bands, PHI1 (25--43$\umu$m) and PHI2 (47-80$\umu$m), and the 4 PPI channels (96--235$\umu$m). Blue dashed line shows the limiting flux density, as defined in Sec. \ref{sec:limiting_flux}, reached by \texttt{XID+} with flat flux priors and the Wiener-filtered detection prior catalogue. The dash-dotted pink line shows the results from \texttt{XID+} with the Deep prior catalogue and flux priors with $\sigma_{\text{s,prior}} = S_{\text{true}}$. 
    The orange solid line shows the classical confusion limits from B24 and the red triangles show the 5$\sigma$ baseline point source sensitives in each of the channels. Also plotted are two model SEDs of galaxies from \protect\cite{Kirkpatrick2015} at z=2 with luminosity L = $10^{12.3} L_{\odot}$, corresponding to the knee of the FIR luminosity function at this redshift \citep{Magnelli2013}, and fraction of luminosity from AGN emission, \textit{f$_{\text{AGN}}$}, of 0 and 0.4, shown by the dotted grey line and the solid gray line respectively.  \textbf{Bottom:} Limiting flux density reached by \texttt{XID+} relative to the 5$\sigma$ confusion limits for the 10 reddest channels which are confusion-dominated. 
    }
    \label{fig:xid_fluxes}
\end{figure*} 

Proceeding with the Wiener-channeled prior catalogue, we ran \texttt{XID+} was run on each of the 16 noisy PRIMAger maps independently with flat flux priors. Note that the maps used as data input to \texttt{XID+}  are not filtered in any way, they are simply the simulated maps containing both confusion and instrumental noise (as described in Section \ref{sec:primager-noise}). The output from \texttt{XID+} is the full posterior distribution for the flux  in the  channel corresponding to the map of each source in the prior catalogue, including the correlation between sources. The measured flux of a given source for a particular channel is quoted as the median of its marginalised posterior flux distribution.


Figure \ref{fig:mad_vs_true} shows the scaled MAD, $\sigma_{\text{MAD}}$ (defined by equation(\ref{eq:mad})), of the measured fluxes from \texttt{XID+} compared to true fluxes for all true source flux bins for the 10 reddest PRIMAger channels. It also shows the chosen `limiting flux' threshold at $\sigma_{\text{MAD}} = 0.2$, corresponding to a measured flux accuracy of 20 per cent  ($5\sigma$), as defined in equation (\ref{eq:limiting_flux}). 

The limiting fluxes reached by \texttt{XID+} in each of the 16 noisy PRIMAger maps are shown in Figure \ref{fig:xid_fluxes} by the dashed blue line. These are compared to the classical confusion limits for each map as calculated by B24 (solid orange line, also given in Table \ref{tab:filter_stats}). For all six PHI1 maps, which are limited by the instrumental noise rather than the confusion noise, \texttt{XID+} is able to accurately measure source fluxes down to within a factor of 1.35 of the $5\sigma$ instrumental noise. For the remaining, redder maps which are confusion-dominated (bands PHI2 and PPI), \texttt{XID+} reaches a limiting flux below the classical confusion limit in each channel. As the bottom panel of Figure \ref{fig:xid_fluxes} shows, the gain in depth relative to the classical confusion limit steadily increases through the representative channels of band PHI2. Starting at the PHI2\_1 channel, accurate fluxes are recovered down to the confusion limit of this channel. By PHI2\_6, fluxes which are a factor of $\sim$ 2 below the respective confusion limit are accurately recovered. For the 4 PPI channels (96-235$\umu$m), this is improved to a factor of $\sim$ 3.

These results are also compared to two galaxy SED models from \cite{Kirkpatrick2015}. One is a star-forming galaxy template at $z=2$ and a luminosity of $L = 10^{12.3} L_{\odot}$ with no AGN emission contributing to the total IR luminosity (\textit{f$_{\text{AGN}}$} $= 0$), shown by the dotted grey line in Figure \ref{fig:xid_fluxes}. The other SED template, however, has \textit{f$_{\text{AGN}}$} $= 0.4$\footnote{$f_{\rm AGN}$ is the ratio of the AGN luminosity to the sum of the AGN and dust luminosities, $f_{\rm AGN}$ = L$_{\rm AGN}$ / (L$_{\rm dust}$ + L$_{\rm AGN}$). In this case the AGN SED is a  type-2 AGN.}
and is shown by the solid grey line. Distinguishing between these two types of objects is important to the extragalactic science case for PRIMA. Being able to do so enables the study of the impact that AGN have on galaxy evolution. The limiting flux results from \texttt{XID+} show that accurate fluxes can be obtained for both objects up to $\leq 100\umu$m,  spanning a range where there is significant distinction between these two SEDs.

\subsection{XID+ Photometry with Deep Prior Catalogue}
\label{sec:xid_deep_results}

Figure \ref{fig:xid_fluxes} also shows the results from the \texttt{XID+} run with the Deep prior source catalogue with flux prior knowledge of $\sigma_{\text{prior}}/S_{\text{true}} = 1$ (pink dash-dotted line). Utilising these more informative priors allows for significantly deeper limiting fluxes to be reached, particularly for the PPI1-PPI3 channels where the limiting flux is more than an order of magnitude below the classical confusion limits. Additionally, for the PHI1 band and blue PHI2 channels, the limiting flux is pushed down to the instrumental noise of the simulated survey.

Comparing again against the two model SED templates with differing \textit{f$_{\text{AGN}}$}, these deeper limiting fluxes allow for these two objects to be accurately observed in the two reddest PRIMAger channels (PPI3 and PPI4).

\section{Discussion}
\label{sec:discussions}
\subsection{Alternative methods}
In this paper we have focused on quantifying how much fainter than the n\"aive, classical confusion we can probe with PRIMAger using modern, but relatively well-established techniques.  However, it is important to note that the hyperspectral capabilities of PRIMAger will lend themselves well to more sophisticated techniques which are likely to do better. The rich spectral information available in PRIMAger (including the continuous linear variable filters in PH1 and PH2) can augment the spatial information. In this paper we have concentrated on using the high resolution at short wavelengths to provide positional priors at longer wavelengths.  However, this does not exploit the fact that different types of galaxies and galaxies at different redhshifts have different spectral signatures. Simultaneously modelling the spatial and spectral information, which is possible in the \texttt{XID+} framework, would improve these results.  Even a relatively simple stepwise approach of stepping through the channels one-by-one and using the short wavelengths to inform the flux priors of the longer wavelengths would yield benefits (e.g. Wang et al. 2024, sub. and see Section~\ref{sec:prior_flux_info_disucssion}) . Furthermore, sophisticated tools are being developed rapidly in the context of AI and machine learning  and we note in Appendix \ref{sec:deconvolution_alternative} that impressive deconvolution results are not limited to this prior-based deblending technique, but are a general property of the hyperspectral imaging dataset.  

PRIMager will also be working alongside spectral imaging capabilities from the FIRESS instrument. Tools like CIGALE \citep{Boquien2019} can model simultaneously PRIMAger photometry and FIRESS spectral data allowing us to consistently model dust and gas.

It is also worth noting that our investigation has been restricted to deep surveys, future work is needed to assess wide surveys. As shown in B24 the wide fields surveys will also be affected by confusion, albeit to a less extent and being confusion limited at longer wavelengths than the deep surveys.  For the wide surveys the reduced sensitivity at shorter wavelengths will have an impact on the prior catalogues that can be self-consistently constructed from PRIMA data, and hence the deblending performance. This can also be addressed by using multi-band techniques in the detection process, e.g. generalising the Wiener filtering to multi-bands.

\subsection{Properties of Galaxies Accessible to PRIMAger}


Having quantified how accurately we can measure fluxes and hence the flux limits at which we can accurately determine fluxes (i.e. to ``detect'' galaxies) it is important to consider the implications for studies of galaxy properties. It is thus instructive to consider the detectability of galaxies in physical parameter space.

\subsubsection{Redshift, SFR Plane}

\begin{figure*}
\centering
\begin{tabular}{cc}
\includegraphics[width=7.5cm]{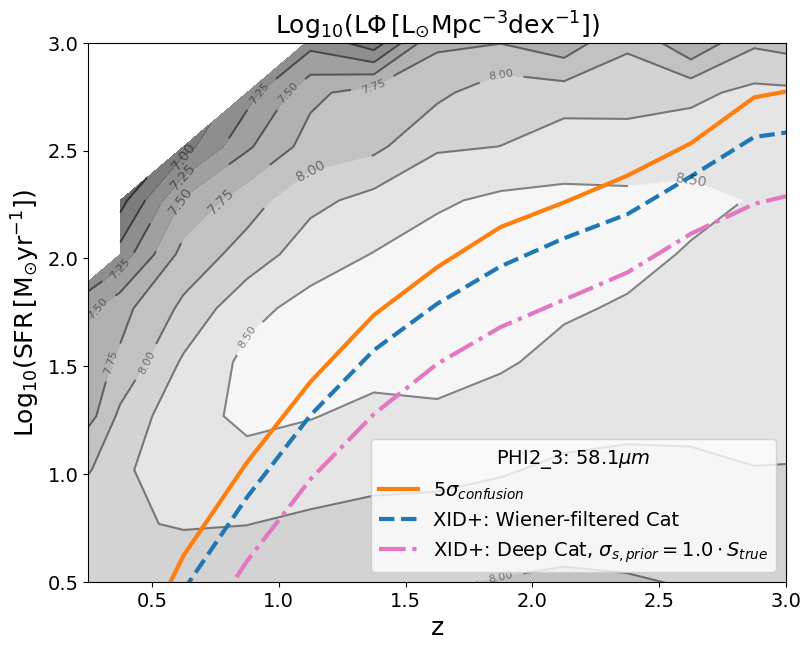} & 
\includegraphics[width=7.5cm]{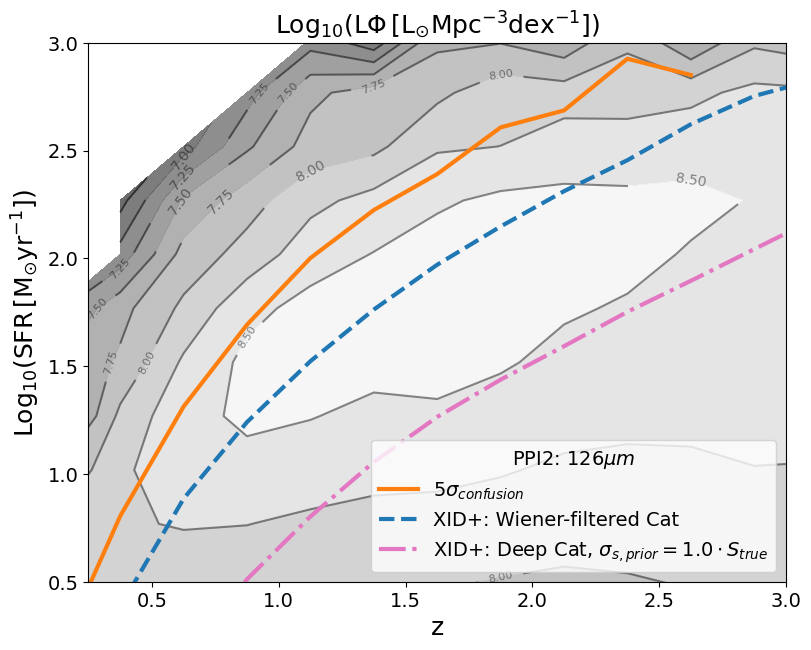}  
\end{tabular}
\caption{\label{fig:z-sfr} Regions of the redshift--SFR plane accessible to PRIMAger at the given limiting fluxes reached by \texttt{XID+} in the PHI2\_3 (left) and PPI2 (right) channels. 
Sources which lie within above the lines are those whose fluxes  can be accurately measured in the given channels using \texttt{XID+}. Additionally, the FIR luminosity density as a function of redshift and SFR is shown by the grey-scale contours. The classical confusion limits from B24 in each channel are shown by the solid orange lines. The limiting fluxes from \texttt{XID+} with the Wiener-filtered prior catalogue and the Deep prior catalogue with flux priors of $\sigma_{\text{s,prior}} = S_{\text{true}}$ are shown by the dashed blue lines and the dash-dotted pink lines, respectively. }
\end{figure*}
We firstly consider the detectability of the significant star formation (as traced by the FIR luminosity density) as a function of SFR and redshift.
The underlying FIR luminosity  density of the SIDES simulation, $L\phi$, as a function of SFR and redshift is indicated by the grey-scale contours in Figure \ref{fig:z-sfr}.
For each channel we translate from a limiting flux from \texttt{XID+} to a limiting SFR as follows.
We select all sources from the SIDES simulation whose true fluxes are within 10 percent of the limiting flux. The limiting SFR is defined as the median SFR of these sources. 

These are shown in Figure \ref{fig:z-sfr} for the limiting fluxes from the two \texttt{XID+} runs with the Wiener-filtered and Deep prior catalogues by the blue dashed and pink dash-dotted lines, respectively, for two of the channels. The region of the z-SFR plane above these lines are where sources have fluxes in the given channel which can be accurately measured for the given method. 

The limiting boundary due to the confusion limit is also estimated in the same way (solid orange lines). As can be seen in the right panel of Figure \ref{fig:z-sfr}, the confusion limit in the PPI2 channel prohibits sources which form the peak of the luminosity density from being reliably recovered. Utilising \texttt{XID+} allows for this peak to begin to be probed even with the low source density prior catalogue and no flux prior information. With the more extensive prior catalogue with additional prior flux information, the full peak of the luminosity density can be explored.

\subsubsection{Stellar mass, SFR Plane}

\begin{figure*}
\centering
\begin{tabular}{c}
\includegraphics[width=\linewidth]{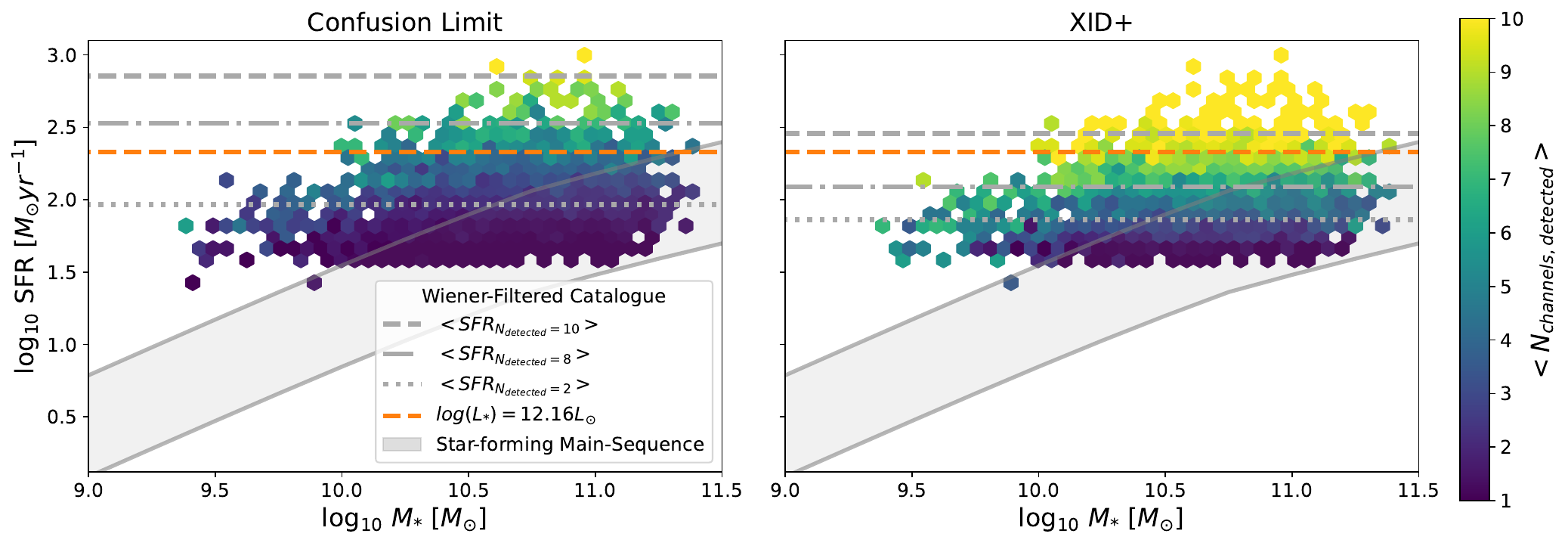} \\
\includegraphics[width=\linewidth]{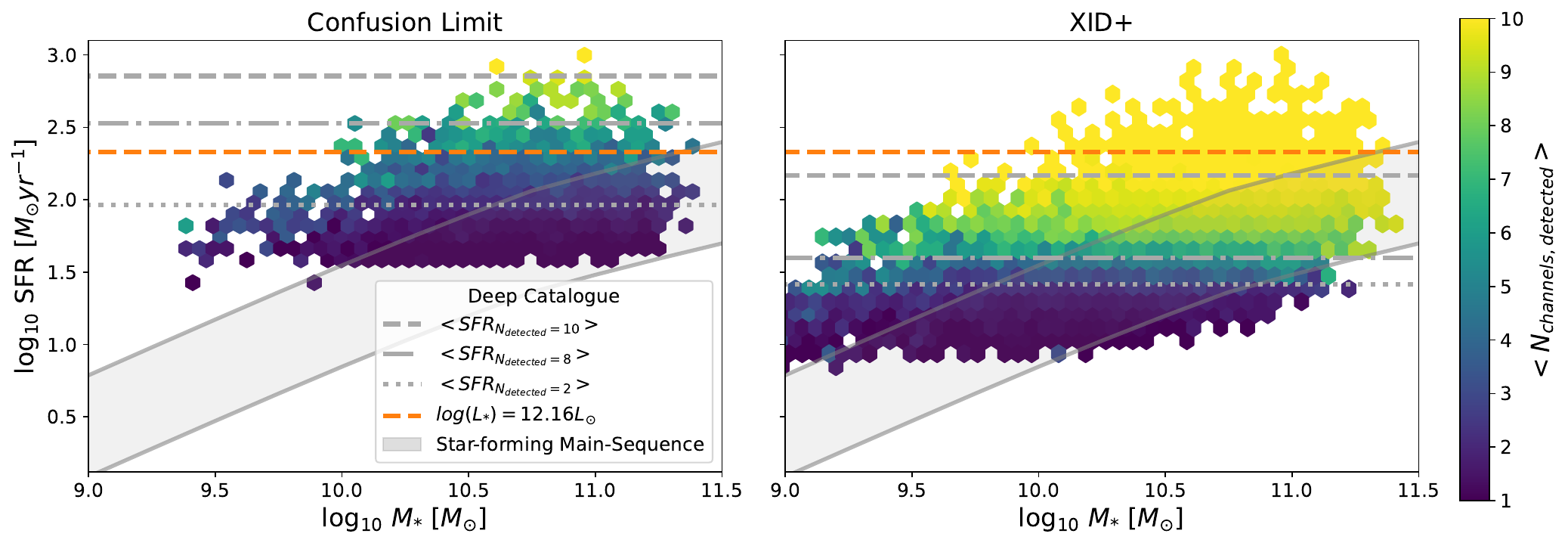}
\end{tabular}
\caption{\label{fig:ms_recovered} Locations of galaxies recoverable by \texttt{XID+} in the stellar mass--SFR plane  for a single redshift bin of $1.5 < z \leq 2.0$, colour-coded by the number of bands in which they could be detected.  A galaxy is considered detectable if it has a true flux above the limiting flux in at least one channel for the confusion-dominated maps (PHI2\_1-PPI4; 47--235$\umu$m). The limiting fluxes are taken as the classical confusion limits from B24 for the top left and bottom left panels and as the limiting flux results from \texttt{XID+} runs with the Wiener-filtered and Deep prior catalogues for the top right and bottom right panels, respectively. Colour-scale shows the average number of channels which the sources within each M$_{*}$-SFR bin are detected in. Overplotted are the average SFR values for all sources which are detected in 2, 8 and 10 of the PHI2\_1-PPI4 channels, shown by the dotted, dash-dotted and dashed grey horizontal lines, respectively. The orange dashed horizontal line shows the SFR of the knee of the FIR luminosity function at $z = 1.75$ from \protect\cite{Magnelli2013}.  The star-forming main-sequence curves from the literature are indicated by the shaded region. This shows that \texttt{XID+} can recover multi-band photometry into the main sequence with galaxies detected self-consistently by PRIMA (top right) and substantially spanning the main-sequence with deeper (external) prior catalogues (bottom right).}
\end{figure*}

We can consider that a source is recovered if it has a true flux in at least one channel for the confusion-dominated maps (PHI2\_1-PPI4; 47--235$\umu$m) which is above the corresponding limiting flux in that channel from \texttt{XID+}. Revisiting the stellar mass-SFR plane for a single redshift bin of $1.5 < z \leq 2.0$ for the sources in the two prior catalogues used for the two \texttt{XID+} runs (originally shown by the left and right panels in Figure \ref{fig:MS_cats}), we can identify which of these sources are recovered. The right-hand panels in Figure \ref{fig:ms_recovered} show the recovered sources meeting the above criteria from the two \texttt{XID+} runs relative to the star-forming main-sequence (MS) from the literature. The left-hand panels show the sources which are recovered above the classical confusion limits determined by B24. For each M$_{*}$-SFR bin, the average number of channels in which the sources within that bin are recovered is also calculated and shown by the colour-scale. Additionally, an average SFR value for all sources which are detected in 2, 8 and 10 of the PHI2\_1-PPI4 channels are also shown. The latter two ensure that at least two detections are made in the 96--235$\umu$m channels and therefore robustly recovering a given galaxy in the FIR regime. 

For the Wiener-filtered prior catalogue (top panels), \texttt{XID+} recovers a comparable number of sources for this prior catalogue and redshift bin as those recovered above the classical confusion limits. This is due to the majority of the sources being detected in the shortest wavelength channel considered for the selection, PHI2\_1 (47$\umu$m), where the limiting flux from this run of \texttt{XID+} is comparable to the classical confusion limit. However, \texttt{XID+} is able to recover these sources in more channels. As such, it is able to accurately sample the FIR regime of galaxy SEDs (by detecting the galaxy and measuring its flux to an accuracy of better than 20 per cent in 8--10 channels in PHI2 and PPI bands) down to log$_{10}$(SFR) $\sim$2--2.5 $M_{\odot} {\rm yr}^{-1}$. This provides an improvement of 0.5 dex compared to what can be recovered above the classical confusion limits. Also shown is the SFR of the knee of the FIR luminosity function at $z = 1.75$ from \cite{Magnelli2013}, which has log$(L_{*}) = 12.16 ~L_{\odot}$ and a corresponding log(SFR) $= 2.33 ~M_{\odot} {\rm yr}^{-1}$, using the conversion from \cite{Kennicutt2012}. \texttt{XID+} is able to recover this knee of the FIR luminosity function in at least 8 of the 10 channels covering 47--235~$\umu$m.  

For the Deep prior catalogue run of \texttt{XID+} utilising weak prior flux information, this is improved further, detecting sources in 8--10 channels down to log$_{10}$(SFR) $\sim$1.6--2.1, which is $\sim$ an order of magnitude below what is reached for sources above the classical confusion limits. Moreover, this run of \texttt{XID+} is able to recover $\sim$3.6x more sources than those above the classical confusion limits for this prior source catalogue and redshift bin.

\subsection{Obtaining Prior Flux Information}
\label{sec:prior_flux_info_disucssion}

For the runs of \texttt{XID+} which have included prior flux knowledge, we have employed a toy model to represent the constraining power of the prior flux knowledge. In reality, these flux priors would  need to be obtained via some modelling of the SED of the source. When PRIMA is launched there will be a wealth of deep ancillary photometry available from contempoary missions and ground-based facilities. SED-fitting of this data could be performed to estimate the flux of the source in the particular channel map which requires de-blending. The choice of SED modelling procedure as well as the type of ancillary data available to fit (e.g. radio and MIR photometry vs. only UV/optical) will inevitably impact the accuracy of the modelling \citep{Pacifici2023, Thorne2023}. Moreover, even if the flux prior information is not constraining for specific sources and is only representative of typical galaxy populations, modelling these will still reduce the confusion noise. Exploiting prior flux information would still allow atypical galaxies to be detected by looking for cases where the posterior significantly departs from the prior distribution or where the model does not fit the data well, through posterior predictive checking. 

In addition to utilising ancillary photometry, another approach which only requires data from PRIMAger is possible due to the probe's extensive spectral coverage and resolution. Source detection and photometry can be performed on the shorter-wavelength band PHI1 maps (i.e. $\lambda < 40 \umu$m), which are not confusion-dominated. Probabilistic SED fitting using an SED library (e.g. from CIGALE) can then be performed on these extracted fluxes to estimate, for example, the fluxes of the sources in the next 3 channels (in ascending wavelength order), providing the flux priors to be used to de-blend the corresponding maps. This step-wise method can be repeated so that the reddest and most confused maps will have prior flux information which is determined from the shorter wavelength maps. Applying this step-wise method and determining whether it can provide prior flux knowledge which is sufficiently informative is beyond the scope of this paper, but we outline it here to be tested in the context of PRIMAger in future work.



\section{Conclusions}
\label{sec:conclusions}

In this work, we have shown that confusion mitigation methods utilising positional priors successfully demonstrated on {\em Herschel} datasets will allow PRIMAger to reliably extract fluxes well below the classical confusion limits.


We have tested these mitigation methods on mock data that simulate a 1500~hr~deg$^{-2}$ depth hyperspectal imaging survey with PRIMager, from 25--235$\umu$m, using a  sky,  observatory and instrument model that provides maps with realistic confusion and ``instrumental'' noise.

We have demonstrated that we can produce catalogues of galaxies with high purity from the PRIMager images alone (i.e. blindly) using a Wiener-filter optimised to suppress both forms of noise. Specifically, we have produced catalogues with 95 per cent purity reaching 55--117 $\umu$Jy in S$_\mathrm{obs}$ in the six PHI1 bands, where the majority of sources are first detected. This blind catalog also reaches a completeness of $\sim$ 83 per cent on sources with S$_\mathrm{True} > \sigma_\mathrm{total}$ in PHI1 bands, with a source density 42k deg$^{-2}$ (or $\sim 0.5$ sources per beam in PPI1 band).


We have then shown that we are able to accurately recover the fluxes of these high purity PRIMager sources from 25--235$\umu$m with no  prior flux information using the Bayesian probabilistic de-blending code \texttt{XID+}. 
We demonstrated  that flux accuracy within 20 per cent of the true flux values are obtained below the confusion limits for all the confusion-dominated maps. A gain of a factor of $\sim$2 below the classical confusion limits (as estimated by B24) is achieved between 72--96~$ \umu$m, as shown in Figures~\ref {fig:mad_vs_true} and \ref{fig:xid_fluxes}. This increases to a factor of $\sim$3 for 126--235~$\umu$m (the reddest channels in the PHI2 band). This allows PRIMAger to recover SEDs out to $\lambda=126$ $~\umu$m for sources at the knee of the infrared luminosity function for $z=2$, as shown in Figure \ref{fig:xid_fluxes}. 

We have also shown that even greater improvements are possible with the introduction of additional prior information, e.g. arising from the detection in, and spectral energy distribution modelling of, other wavelengths with data from other contemporary observatories. We have investigated the impact of increasing the source density of the prior position catalogue alongside varying prior flux knowledge on the flux modelling accuracy of \texttt{XID+}. De-blending of sources at high densities ($>1$ source per beam), or equivalently lower fluxes, benefits significantly from adding prior flux  information to \texttt{XID+}. We show that with  weak prior flux information (a Gaussian prior with dispersion equal to the flux) accurate fluxes for sources are recovered at $\lambda < 80\umu$m down to the instrumental noise level of the survey. This same catalog and flux prior results in recovering fluxes about an order of magnitude below the classical confusion limit at  96--172$\umu$m, and a factor of 6 below the classical confusion confusion at 235$\umu$m.

We have also shown that de-blending with \texttt{XID+} allows a survey such as the one described for PRIMAger to detect and measure accurately  source fluxes for galaxies which contribute to the bulk of the IR luminosity density. Additionally, we have demonstrated that \texttt{XID+} is able to sample the FIR regime of galaxy SEDs with accurate flux measurements in 8--10 of the 10 channels covering 47--235$\umu$m for sources with log(SFR) $\sim$ 2--2.5 at $1.5 \leq z \leq 2.0$. This improves upon what can be achieved above the classical confusion limits by 0.5 dex, as shown in the top panel of Figure \ref{fig:ms_recovered}. Most importantly, these observations are self-contained as the prior source catalogues can be obtained from the shorter wavelength PRIMAger maps, where confusion noise is not dominant, and are subsequently used to de-blend the longer wavelength maps and accurately measure source fluxes. 

We have therefore demonstrated that imaging data from PRIMAger will not be limited by n\"aive, classical confusion noise if deblending with \texttt{XID+} is employed. Accurate flux measurements below the confusion limits are therefore currently achievable using data from PRIMAger in a self-contained way. 

Further improvement can also be achieved both by utilising ancillary data to provide additional prior source positions and prior flux information, and also, with PRIMAger data along by using shorter wavelength data to provide improved priors by utilising \texttt{XID+} in a step-wise process as described in Section \ref{sec:prior_flux_info_disucssion}.


\section*{Contributions}

The contributions of the authors using the Contributor Roles Taxonomy (CRediT) were as follows.
{\bf James Donnellan:} Methodology, Software, Validation, Investigation, Writing - original Draft
{\bf Seb Oliver:} Conceptualization, Supervision, Project Administration, Writing - Review \& Editing, Funding acquisition
{\bf Matthieu Bethermin:}
Resources, Conceptualization, Writing - Review \& Editing
{\bf Longji Bing:} Methodology, Software, Validation, Investigation
{\bf Alberto Bolatto:}
Conceptualization, Writing - Review \& Editing
{\bf Charles M. Bradford:}
Conceptualization, Writing - Review \& Editing
{\bf Denis Burgarella:}
Writing - Review \& Editing
{\bf Laure Ciesla:}
Conceptualization, Writing - Review \& Editing
{\bf Jason Glenn:}
Conceptualization, Writing - Review \& Editing
{\bf Alex Pope:}
Conceptualization, Writing - Review \& Editing
{\bf Stephen Serjeant:} Autoencoder deconvolution, writing - Review \& Editing
{\bf Raphael Shirley:}
Writing - Review \& Editing
{\bf JD T. Smith:}
Conceptualization, Writing - Review \& Editing
{\bf Chris Sorrell:} Autoencoder deconvolution, writing - Review \& Editing

\section*{Acknowledgements}
James Donnellan and Chris Sorrell were supported by the Science and Technology Facilities Council, grant number ST/W006839/1, through the DISCnet Centre for Doctoral Training.  
Seb Oliver acknowledges support from the UK Space Agency through ST/Y006038/1.
Longji Bing acknowledges funding from STFC through ST/T000473/1 and ST/X001040/1.

This paper makes use of \texttt{ASTROPY} \citep{astropy} a community developed core \texttt{PYTHON} package for Astronomy.

\section*{Data Availability}
All data are available upon request.



\bibliographystyle{mnras}
\bibliography{primager_confusion} 

\begin{thebibliography}{}
\makeatletter
\relax
\def\mn@urlcharsother{\let\do\@makeother \do\$\do\&\do\#\do\^\do\_\do\%\do\~}
\def\mn@doi{\begingroup\mn@urlcharsother \@ifnextchar [ {\mn@doi@} {\mn@doi@[]}}
\def\mn@doi@[#1]#2{\def\@tempa{#1}\ifx\@tempa\@empty \href {http://dx.doi.org/#2} {doi:#2}\else \href {http://dx.doi.org/#2} {#1}\fi \endgroup}
\def\mn@eprint#1#2{\mn@eprint@#1:#2::\@nil}
\def\mn@eprint@arXiv#1{\href {http://arxiv.org/abs/#1} {{\tt arXiv:#1}}}
\def\mn@eprint@dblp#1{\href {http://dblp.uni-trier.de/rec/bibtex/#1.xml} {dblp:#1}}
\def\mn@eprint@#1:#2:#3:#4\@nil{\def\@tempa {#1}\def\@tempb {#2}\def\@tempc {#3}\ifx \@tempc \@empty \let \@tempc \@tempb \let \@tempb \@tempa \fi \ifx \@tempb \@empty \def\@tempb {arXiv}\fi \@ifundefined {mn@eprint@\@tempb}{\@tempb:\@tempc}{\expandafter \expandafter \csname mn@eprint@\@tempb\endcsname \expandafter{\@tempc}}}

\bibitem[\protect\citeauthoryear{{Astropy Collaboration} et~al.,}{{Astropy Collaboration} et~al.}{2013}]{astropy}
{Astropy Collaboration} et~al., 2013, \mn@doi [\aap] {10.1051/0004-6361/201322068}, \href {https://ui.adsabs.harvard.edu/abs/2013A&A...558A..33A} {558, A33}

\bibitem[\protect\citeauthoryear{{B{\'e}thermin} et~al.,}{{B{\'e}thermin} et~al.}{2017}]{Sides1}
{B{\'e}thermin} M.,  et~al., 2017, \mn@doi [\aap] {10.1051/0004-6361/201730866}, \href {https://ui.adsabs.harvard.edu/abs/2017A&A...607A..89B} {607, A89}

\bibitem[\protect\citeauthoryear{{B{\'e}thermin} et~al.,}{{B{\'e}thermin} et~al.}{2022}]{Sides2}
{B{\'e}thermin} M.,  et~al., 2022, \mn@doi [\aap] {10.1051/0004-6361/202243888}, \href {https://ui.adsabs.harvard.edu/abs/2022A&A...667A.156B} {667, A156}

\bibitem[\protect\citeauthoryear{{Boquien}, {Burgarella}, {Roehlly}, {Buat}, {Ciesla}, {Corre}, {Inoue}  \& {Salas}}{{Boquien} et~al.}{2019}]{Boquien2019}
{Boquien} M.,  {Burgarella} D.,  {Roehlly} Y.,  {Buat} V.,  {Ciesla} L.,  {Corre} D.,  {Inoue} A.~K.,   {Salas} H.,  2019, \mn@doi [\aap] {10.1051/0004-6361/201834156}, \href {https://ui.adsabs.harvard.edu/abs/2019A&A...622A.103B} {622, A103}

\bibitem[\protect\citeauthoryear{{Burgarella} et~al.,}{{Burgarella} et~al.}{2013a}]{Burgarella2013}
{Burgarella} D.,  et~al., 2013a, \mn@doi [\aap] {10.1051/0004-6361/201321651}, \href {https://ui.adsabs.harvard.edu/abs/2013A&A...554A..70B} {554, A70}

\bibitem[\protect\citeauthoryear{{Burgarella} et~al.,}{{Burgarella} et~al.}{2013b}]{2013A&A...554A..70B}
{Burgarella} D.,  et~al., 2013b, \mn@doi [\aap] {10.1051/0004-6361/201321651}, \href {https://ui.adsabs.harvard.edu/abs/2013A&A...554A..70B} {554, A70}

\bibitem[\protect\citeauthoryear{{Calzetti}, {Armus}, {Bohlin}, {Kinney}, {Koornneef}  \& {Storchi-Bergmann}}{{Calzetti} et~al.}{2000}]{Calzetti2000}
{Calzetti} D.,  {Armus} L.,  {Bohlin} R.~C.,  {Kinney} A.~L.,  {Koornneef} J.,   {Storchi-Bergmann} T.,  2000, \mn@doi [\apj] {10.1086/308692}, \href {https://ui.adsabs.harvard.edu/abs/2000ApJ...533..682C} {533, 682}

\bibitem[\protect\citeauthoryear{{Cardelli}, {Clayton}  \& {Mathis}}{{Cardelli} et~al.}{1989}]{Cardelli1989}
{Cardelli} J.~A.,  {Clayton} G.~C.,   {Mathis} J.~S.,  1989, \mn@doi [\apj] {10.1086/167900}, \href {https://ui.adsabs.harvard.edu/abs/1989ApJ...345..245C} {345, 245}

\bibitem[\protect\citeauthoryear{{Casey}, {Narayanan}  \& {Cooray}}{{Casey} et~al.}{2014}]{Casey2014}
{Casey} C.~M.,  {Narayanan} D.,   {Cooray} A.,  2014, \mn@doi [\physrep] {10.1016/j.physrep.2014.02.009}, \href {https://ui.adsabs.harvard.edu/abs/2014PhR...541...45C} {541, 45}

\bibitem[\protect\citeauthoryear{{Chapin} et~al.,}{{Chapin} et~al.}{2011}]{Chapin2011}
{Chapin} E.~L.,  et~al., 2011, \mn@doi [\mnras] {10.1111/j.1365-2966.2010.17697.x}, \href {https://ui.adsabs.harvard.edu/abs/2011MNRAS.411..505C} {411, 505}

\bibitem[\protect\citeauthoryear{{Condon}}{{Condon}}{1974}]{Condon}
{Condon} J.~J.,  1974, \mn@doi [\apj] {10.1086/152714}, \href {https://ui.adsabs.harvard.edu/abs/1974ApJ...188..279C} {188, 279}

\bibitem[\protect\citeauthoryear{{Dole} et~al.,}{{Dole} et~al.}{2004}]{Dole2004}
{Dole} H.,  et~al., 2004, \mn@doi [\apjs] {10.1086/422690}, \href {https://ui.adsabs.harvard.edu/abs/2004ApJS..154...93D} {154, 93}

\bibitem[\protect\citeauthoryear{{Dole} et~al.,}{{Dole} et~al.}{2006}]{Dole2006}
{Dole} H.,  et~al., 2006, \mn@doi [\aap] {10.1051/0004-6361:20054446}, \href {https://ui.adsabs.harvard.edu/abs/2006A&A...451..417D} {451, 417}

\bibitem[\protect\citeauthoryear{{Duivenvoorden} et~al.,}{{Duivenvoorden} et~al.}{2020}]{Duivenvoorden2020}
{Duivenvoorden} S.,  et~al., 2020, \mn@doi [\mnras] {10.1093/mnras/stz3110}, \href {https://ui.adsabs.harvard.edu/abs/2020MNRAS.491.1355D} {491, 1355}

\bibitem[\protect\citeauthoryear{{Elbaz} et~al.,}{{Elbaz} et~al.}{2011}]{Elbaz2011}
{Elbaz} D.,  et~al., 2011, \mn@doi [\aap] {10.1051/0004-6361/201117239}, \href {https://ui.adsabs.harvard.edu/abs/2011A&A...533A.119E} {533, A119}

\bibitem[\protect\citeauthoryear{{Geach} et~al.,}{{Geach} et~al.}{2017}]{Geach2017}
{Geach} J.~E.,  et~al., 2017, \mn@doi [\mnras] {10.1093/mnras/stw2721}, \href {https://ui.adsabs.harvard.edu/abs/2017MNRAS.465.1789G} {465, 1789}

\bibitem[\protect\citeauthoryear{{Genzel} \& {Cesarsky}}{{Genzel} \& {Cesarsky}}{2000}]{Genzel2000}
{Genzel} R.,  {Cesarsky} C.~J.,  2000, \mn@doi [\araa] {10.1146/annurev.astro.38.1.761}, \href {https://ui.adsabs.harvard.edu/abs/2000ARA&A..38..761G} {38, 761}

\bibitem[\protect\citeauthoryear{{Gruppioni} et~al.,}{{Gruppioni} et~al.}{2010}]{Gruppioni2010}
{Gruppioni} C.,  et~al., 2010, \mn@doi [\aap] {10.1051/0004-6361/201014608}, \href {https://ui.adsabs.harvard.edu/abs/2010A&A...518L..27G} {518, L27}

\bibitem[\protect\citeauthoryear{{Gruppioni} et~al.,}{{Gruppioni} et~al.}{2013a}]{2013MNRAS.432...23G}
{Gruppioni} C.,  et~al., 2013a, \mn@doi [\mnras] {10.1093/mnras/stt308}, \href {https://ui.adsabs.harvard.edu/abs/2013MNRAS.432...23G} {432, 23}

\bibitem[\protect\citeauthoryear{{Gruppioni} et~al.,}{{Gruppioni} et~al.}{2013b}]{Gruppioni2013}
{Gruppioni} C.,  et~al., 2013b, \mn@doi [\mnras] {10.1093/mnras/stt308}, \href {https://ui.adsabs.harvard.edu/abs/2013MNRAS.432...23G} {432, 23}

\bibitem[\protect\citeauthoryear{{Hauser} \& {Dwek}}{{Hauser} \& {Dwek}}{2001}]{Hauser2001}
{Hauser} M.~G.,  {Dwek} E.,  2001, \mn@doi [\araa] {10.1146/annurev.astro.39.1.249}, \href {https://ui.adsabs.harvard.edu/abs/2001ARA&A..39..249H} {39, 249}

\bibitem[\protect\citeauthoryear{{Hurley} et~al.,}{{Hurley} et~al.}{2017}]{Hurley}
{Hurley} P.~D.,  et~al., 2017, \mn@doi [\mnras] {10.1093/mnras/stw2375}, \href {https://ui.adsabs.harvard.edu/abs/2017MNRAS.464..885H} {464, 885}

\bibitem[\protect\citeauthoryear{{Kennicutt} \& {Evans}}{{Kennicutt} \& {Evans}}{2012}]{Kennicutt2012}
{Kennicutt} R.~C.,  {Evans} N.~J.,  2012, \mn@doi [\araa] {10.1146/annurev-astro-081811-125610}, \href {https://ui.adsabs.harvard.edu/abs/2012ARA&A..50..531K} {50, 531}

\bibitem[\protect\citeauthoryear{{Kessler} et~al.,}{{Kessler} et~al.}{1996}]{Kessler1996}
{Kessler} M.~F.,  et~al., 1996, \aap, \href {https://ui.adsabs.harvard.edu/abs/1996A&A...315L..27K} {315, L27}

\bibitem[\protect\citeauthoryear{{Kirkpatrick}, {Pope}, {Sajina}, {Roebuck}, {Yan}, {Armus}, {D{\'\i}az-Santos}  \& {Stierwalt}}{{Kirkpatrick} et~al.}{2015}]{Kirkpatrick2015}
{Kirkpatrick} A.,  {Pope} A.,  {Sajina} A.,  {Roebuck} E.,  {Yan} L.,  {Armus} L.,  {D{\'\i}az-Santos} T.,   {Stierwalt} S.,  2015, \mn@doi [\apj] {10.1088/0004-637X/814/1/9}, \href {https://ui.adsabs.harvard.edu/abs/2015ApJ...814....9K} {814, 9}

\bibitem[\protect\citeauthoryear{{Lauritsen}, {Dickinson}, {Bromley}, {Serjeant}, {Lim}, {Gao}  \& {Wang}}{{Lauritsen} et~al.}{2021}]{Lauritsen2021}
{Lauritsen} L.,  {Dickinson} H.,  {Bromley} J.,  {Serjeant} S.,  {Lim} C.-F.,  {Gao} Z.-K.,   {Wang} W.-H.,  2021, \mn@doi [\mnras] {10.1093/mnras/stab2195}, \href {https://ui.adsabs.harvard.edu/abs/2021MNRAS.507.1546L} {507, 1546}

\bibitem[\protect\citeauthoryear{{Leja} et~al.,}{{Leja} et~al.}{2022}]{Leja2022}
{Leja} J.,  et~al., 2022, \mn@doi [\apj] {10.3847/1538-4357/ac887d}, \href {https://ui.adsabs.harvard.edu/abs/2022ApJ...936..165L} {936, 165}

\bibitem[\protect\citeauthoryear{{Leslie} et~al.,}{{Leslie} et~al.}{2020}]{Leslie2020}
{Leslie} S.~K.,  et~al., 2020, \mn@doi [\apj] {10.3847/1538-4357/aba044}, \href {https://ui.adsabs.harvard.edu/abs/2020ApJ...899...58L} {899, 58}

\bibitem[\protect\citeauthoryear{{Long}, {Casey}, {Lagos}, {Lambrides}, {Zavala}, {Champagne}, {Cooper}  \& {Cooray}}{{Long} et~al.}{2022}]{Long2022}
{Long} A.~S.,  {Casey} C.~M.,  {Lagos} C. d.~P.,  {Lambrides} E.~L.,  {Zavala} J.~A.,  {Champagne} J.,  {Cooper} O.~R.,   {Cooray} A.~R.,  2022, \mn@doi [arXiv e-prints] {10.48550/arXiv.2211.02072}, \href {https://ui.adsabs.harvard.edu/abs/2022arXiv221102072L} {p. arXiv:2211.02072}

\bibitem[\protect\citeauthoryear{{Madau} \& {Dickinson}}{{Madau} \& {Dickinson}}{2014}]{Madau2014}
{Madau} P.,  {Dickinson} M.,  2014, \mn@doi [\araa] {10.1146/annurev-astro-081811-125615}, \href {https://ui.adsabs.harvard.edu/abs/2014ARA&A..52..415M} {52, 415}

\bibitem[\protect\citeauthoryear{{Magnelli} et~al.,}{{Magnelli} et~al.}{2014}]{Magnelli2013}
{Magnelli} B.,  et~al., 2014, \mn@doi [\aap] {10.1051/0004-6361/201322217}, \href {https://ui.adsabs.harvard.edu/abs/2014A&A...561A..86M} {561, A86}

\bibitem[\protect\citeauthoryear{{Ma{\l}ek} et~al.,}{{Ma{\l}ek} et~al.}{2018}]{Malek2018}
{Ma{\l}ek} K.,  et~al., 2018, \mn@doi [\aap] {10.1051/0004-6361/201833131}, \href {https://ui.adsabs.harvard.edu/abs/2018A&A...620A..50M} {620, A50}

\bibitem[\protect\citeauthoryear{{Meurer}, {Heckman}  \& {Calzetti}}{{Meurer} et~al.}{1999}]{Meurer1999}
{Meurer} G.~R.,  {Heckman} T.~M.,   {Calzetti} D.,  1999, \mn@doi [\apj] {10.1086/307523}, \href {https://ui.adsabs.harvard.edu/abs/1999ApJ...521...64M} {521, 64}

\bibitem[\protect\citeauthoryear{{Moullet} et~al.,}{{Moullet} et~al.}{2023}]{Moullet2023}
{Moullet} A.,  et~al., 2023, \mn@doi [arXiv e-prints] {10.48550/arXiv.2310.20572}, \href {https://ui.adsabs.harvard.edu/abs/2023arXiv231020572M} {p. arXiv:2310.20572}

\bibitem[\protect\citeauthoryear{{Narayanan}, {Dav{\'e}}, {Johnson}, {Thompson}, {Conroy}  \& {Geach}}{{Narayanan} et~al.}{2018}]{Narayanan2018}
{Narayanan} D.,  {Dav{\'e}} R.,  {Johnson} B.~D.,  {Thompson} R.,  {Conroy} C.,   {Geach} J.,  2018, \mn@doi [\mnras] {10.1093/mnras/stx2860}, \href {https://ui.adsabs.harvard.edu/abs/2018MNRAS.474.1718N} {474, 1718}

\bibitem[\protect\citeauthoryear{{Neugebauer} et~al.,}{{Neugebauer} et~al.}{1984}]{Neugebauer1984}
{Neugebauer} G.,  et~al., 1984, \mn@doi [\apjl] {10.1086/184209}, \href {https://ui.adsabs.harvard.edu/abs/1984ApJ...278L...1N} {278, L1}

\bibitem[\protect\citeauthoryear{{Nguyen} et~al.,}{{Nguyen} et~al.}{2010}]{Nguyen2010}
{Nguyen} H.~T.,  et~al., 2010, \mn@doi [\aap] {10.1051/0004-6361/201014680}, \href {https://ui.adsabs.harvard.edu/abs/2010A&A...518L...5N} {518, L5}

\bibitem[\protect\citeauthoryear{{Oliver} et~al.,}{{Oliver} et~al.}{2012}]{2012MNRAS.424.1614O}
{Oliver} S.~J.,  et~al., 2012, \mn@doi [\mnras] {10.1111/j.1365-2966.2012.20912.x}, \href {https://ui.adsabs.harvard.edu/abs/2012MNRAS.424.1614O} {424, 1614}

\bibitem[\protect\citeauthoryear{{Pacifici} et~al.,}{{Pacifici} et~al.}{2023}]{Pacifici2023}
{Pacifici} C.,  et~al., 2023, \mn@doi [\apj] {10.3847/1538-4357/acacff}, \href {https://ui.adsabs.harvard.edu/abs/2023ApJ...944..141P} {944, 141}

\bibitem[\protect\citeauthoryear{{Pearson}, {Wang}, {van der Tak}, {Hurley}, {Burgarella}  \& {Oliver}}{{Pearson} et~al.}{2017}]{Pearson2017}
{Pearson} W.~J.,  {Wang} L.,  {van der Tak} F.~F.~S.,  {Hurley} P.~D.,  {Burgarella} D.,   {Oliver} S.~J.,  2017, \mn@doi [\aap] {10.1051/0004-6361/201630105}, \href {https://ui.adsabs.harvard.edu/abs/2017A&A...603A.102P} {603, A102}

\bibitem[\protect\citeauthoryear{{Pearson} et~al.,}{{Pearson} et~al.}{2018}]{Pearson2018}
{Pearson} W.~J.,  et~al., 2018, \mn@doi [\aap] {10.1051/0004-6361/201832821}, \href {https://ui.adsabs.harvard.edu/abs/2018A&A...615A.146P} {615, A146}

\bibitem[\protect\citeauthoryear{{Pilbratt} et~al.,}{{Pilbratt} et~al.}{2010}]{Pilbratt2010}
{Pilbratt} G.~L.,  et~al., 2010, \mn@doi [\aap] {10.1051/0004-6361/201014759}, \href {https://ui.adsabs.harvard.edu/abs/2010A&A...518L...1P} {518, L1}

\bibitem[\protect\citeauthoryear{{Puget}, {Abergel}, {Bernard}, {Boulanger}, {Burton}, {Desert}  \& {Hartmann}}{{Puget} et~al.}{1996}]{Puget1996}
{Puget} J.~L.,  {Abergel} A.,  {Bernard} J.~P.,  {Boulanger} F.,  {Burton} W.~B.,  {Desert} F.~X.,   {Hartmann} D.,  1996, \aap, \href {https://ui.adsabs.harvard.edu/abs/1996A&A...308L...5P} {308, L5}

\bibitem[\protect\citeauthoryear{{Reddy}, {Steidel}, {Pettini}, {Adelberger}, {Shapley}, {Erb}  \& {Dickinson}}{{Reddy} et~al.}{2008}]{Reddy2008}
{Reddy} N.~A.,  {Steidel} C.~C.,  {Pettini} M.,  {Adelberger} K.~L.,  {Shapley} A.~E.,  {Erb} D.~K.,   {Dickinson} M.,  2008, \mn@doi [\apjs] {10.1086/521105}, \href {https://ui.adsabs.harvard.edu/abs/2008ApJS..175...48R} {175, 48}

\bibitem[\protect\citeauthoryear{{Riccio} et~al.,}{{Riccio} et~al.}{2021}]{Riccio2021}
{Riccio} G.,  et~al., 2021, \mn@doi [\aap] {10.1051/0004-6361/202140854}, \href {https://ui.adsabs.harvard.edu/abs/2021A&A...653A.107R} {653, A107}

\bibitem[\protect\citeauthoryear{{Riechers} et~al.,}{{Riechers} et~al.}{2013}]{2013Natur.496..329R}
{Riechers} D.~A.,  et~al., 2013, \mn@doi [\nat] {10.1038/nature12050}, \href {https://ui.adsabs.harvard.edu/abs/2013Natur.496..329R} {496, 329}

\bibitem[\protect\citeauthoryear{{Rowan-Robinson} et~al.,}{{Rowan-Robinson} et~al.}{2018}]{Rowan2018}
{Rowan-Robinson} M.,  et~al., 2018, \mn@doi [\aap] {10.1051/0004-6361/201832671}, \href {https://ui.adsabs.harvard.edu/abs/2018A&A...619A.169R} {619, A169}

\bibitem[\protect\citeauthoryear{{Shim} et~al.,}{{Shim} et~al.}{2023}]{2023AJ....165...31S}
{Shim} H.,  et~al., 2023, \mn@doi [\aj] {10.3847/1538-3881/aca09c}, \href {https://ui.adsabs.harvard.edu/abs/2023AJ....165...31S} {165, 31}

\bibitem[\protect\citeauthoryear{{Shirley} et~al.,}{{Shirley} et~al.}{2019}]{Shirely2019}
{Shirley} R.,  et~al., 2019, \mn@doi [\mnras] {10.1093/mnras/stz2509}, \href {https://ui.adsabs.harvard.edu/abs/2019MNRAS.490..634S} {490, 634}

\bibitem[\protect\citeauthoryear{{Shirley} et~al.,}{{Shirley} et~al.}{2021}]{Shirley2021}
{Shirley} R.,  et~al., 2021, \mn@doi [\mnras] {10.1093/mnras/stab1526}, \href {https://ui.adsabs.harvard.edu/abs/2021MNRAS.507..129S} {507, 129}

\bibitem[\protect\citeauthoryear{{Shivaei} et~al.,}{{Shivaei} et~al.}{2017}]{Shivaei2017}
{Shivaei} I.,  et~al., 2017, \mn@doi [\apj] {10.3847/1538-4357/aa619c}, \href {https://ui.adsabs.harvard.edu/abs/2017ApJ...837..157S} {837, 157}

\bibitem[\protect\citeauthoryear{{Speagle}, {Steinhardt}, {Capak}  \& {Silverman}}{{Speagle} et~al.}{2014}]{Speagle2014}
{Speagle} J.~S.,  {Steinhardt} C.~L.,  {Capak} P.~L.,   {Silverman} J.~D.,  2014, \mn@doi [\apjs] {10.1088/0067-0049/214/2/15}, \href {https://ui.adsabs.harvard.edu/abs/2014ApJS..214...15S} {214, 15}

\bibitem[\protect\citeauthoryear{{Thorne}, {Robotham}, {Bellstedt}  \& {Davies}}{{Thorne} et~al.}{2023}]{Thorne2023}
{Thorne} J.~E.,  {Robotham} A. S.~G.,  {Bellstedt} S.,   {Davies} L. J.~M.,  2023, \mn@doi [\mnras] {10.1093/mnras/stad1361}, \href {https://ui.adsabs.harvard.edu/abs/2023MNRAS.522.6354T} {522, 6354}

\bibitem[\protect\citeauthoryear{{Viero} et~al.,}{{Viero} et~al.}{2015}]{Viero2015}
{Viero} M.~P.,  et~al., 2015, \mn@doi [\apjl] {10.1088/2041-8205/809/2/L22}, \href {https://ui.adsabs.harvard.edu/abs/2015ApJ...809L..22V} {809, L22}

\bibitem[\protect\citeauthoryear{{Wang} et~al.,}{{Wang} et~al.}{2021}]{Wang2021}
{Wang} L.,  et~al., 2021, \mn@doi [\aap] {10.1051/0004-6361/202038811}, \href {https://ui.adsabs.harvard.edu/abs/2021A&A...648A...8W} {648, A8}

\bibitem[\protect\citeauthoryear{{Werner} et~al.,}{{Werner} et~al.}{2004}]{Werner2004}
{Werner} M.~W.,  et~al., 2004, \mn@doi [\apjs] {10.1086/422992}, \href {https://ui.adsabs.harvard.edu/abs/2004ApJS..154....1W} {154, 1}

\bibitem[\protect\citeauthoryear{{Zavala} et~al.,}{{Zavala} et~al.}{2023}]{2023ApJ...943L...9Z}
{Zavala} J.~A.,  et~al., 2023, \mn@doi [\apjl] {10.3847/2041-8213/acacfe}, \href {https://ui.adsabs.harvard.edu/abs/2023ApJ...943L...9Z} {943, L9}

\makeatother
\end{thebibliography}




\appendix


\section{Choice of Limiting Flux Statistic}
\label{sec:choice_of_statistic}

\begin{figure}
    \centering
    \includegraphics[width=\linewidth]{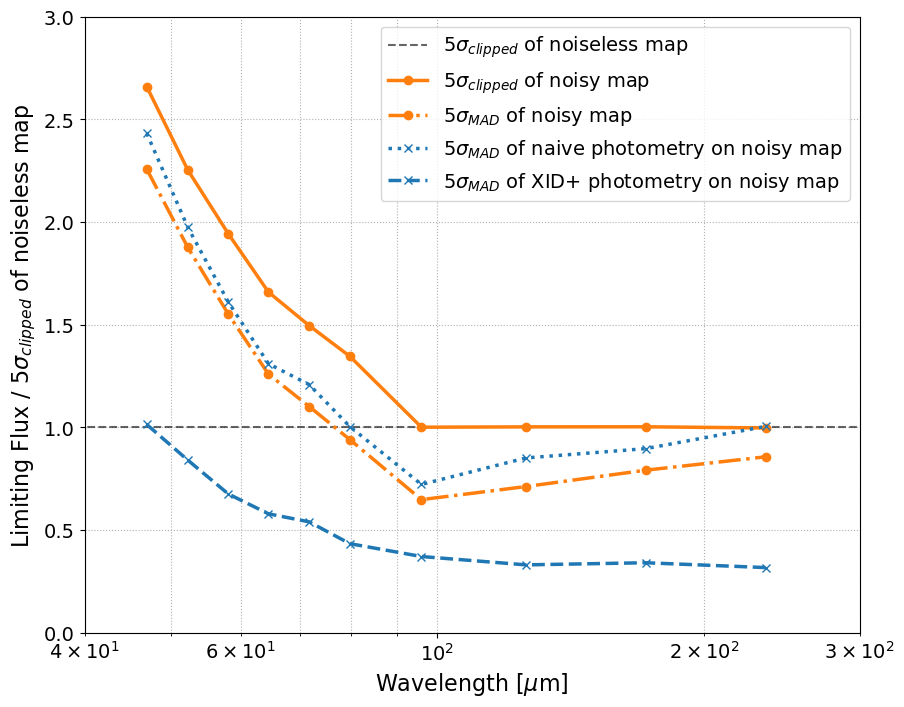}
    \caption{Comparison of different statistics used to measure noise in the simulated maps and the performance of different photometry methods for the 10 reddest PRIMAger filters (47--235$\umu$m).}
    \label{fig:statistic_comparison}
\end{figure}
Since confusion is highly non-Gaussian, any measurement of noise is very sensitive to the statistics being used.
It is thus important to understand the relative contributions to improved performance arising from improved photometry through \texttt{XID+} and through more appropriate statistics. In this Appendix we consider the impact of different statistics in measuring the fluctuations  across the whole map (i.e. for measurements of sky or confusion noise) and then at the location of specific objects (i.e. for point source photometry).

In order to determine whether the limiting flux statistic introduced in Section \ref{sec:limiting_flux} is a reasonable choice, we compare the results from \texttt{XID+} to different measures of noise in the simulated maps. 

Firstly, we apply the same sigma-clipping process used by B24 to determine the classical confusion limits from the noiseless maps (as described in Section \ref{sec:primager-confusion}) to the maps which also contain instrumental noise. This provides an estimate of the total noise in the map and is shown by the solid orange line in Figure \ref{fig:statistic_comparison}. For the PPI maps, the instrumental noise is negligible compared to the confusion noise and therefore the clipped-variance of the noisy map is comparable to that of the noiseless map. For the PHI2 maps, the clipped variance of the noisy map is greater than the confusion limits, as expected because the instrumental noise is non-negligible. This direct measurement of the total noise in the map is also greater than simply adding the two noise components in quadrature as confusion noise is non-Gaussian. 

Secondly, we estimate the MAD of the pixel values as a more  robust estimator of the dispersion of the pixel values in the noisy maps (scaled by a factor of 1.4862, being the ratio between MAD and $\sigma$ for a Gaussian distribution).  This is shown in Figure \ref{fig:statistic_comparison} as the dash-dotted orange line. It is clear that this statistic naturally returns a lower estimate of the noise than the clipped variance. 
This is partially because the MAD statistic ignores both the positive and negative tails, and while the clipping only removes the positive tail, but more importantly because the MAD statistic removes much more of the tails.

We now turn to consider the dispersion metrics at the locations of sources. To investigate this, we conduct a n{\"a}ive photometry to measure the fluxes of the Wiener-filtered catalogue sources in the noisy maps. This involves simply reading  the  value of the map at the position on the source (as the maps are in units of mJy/beam). The MAD-based limiting flux statistic is then applied to these n{\"a}ive photometry measurements. 
The results are shown by the blue dotted line in Figure \ref{fig:statistic_comparison} and are consistent with the MAD measure of the total noise in the maps.
Both, however, are systematically lower than the clipped variance measure of noise ($\sim$ 10-20 per cent lower), implying that some of the gains from \texttt{XID+} compared to the classical confusion limits are due to this choice of the statistic. Despite this, the results from \texttt{XID+} remain below all of the measures of the total noise for all confusion-dominated maps.

\section{Robustness of deconvolution techniques} \label{sec:deconvolution_alternative} 
\begin{figure*}   \centering    \includegraphics[width=\linewidth]{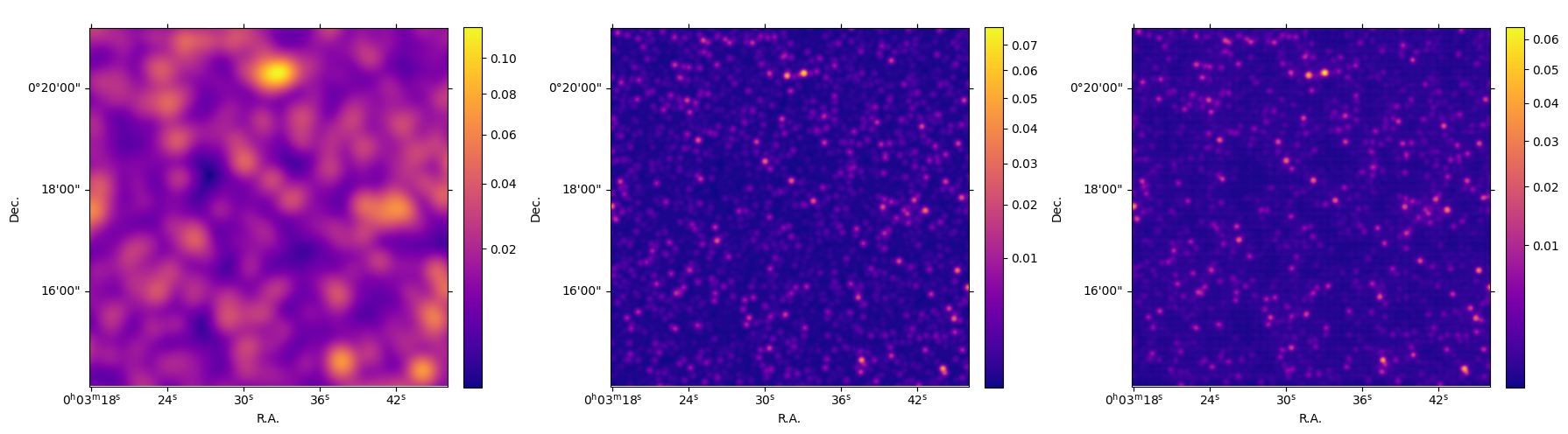}    \caption{This figure shows $424\times424$ arcsec$^2$  postage stamps of the simulated 235$\umu$m PRIMA SIDES image (left) compared to the autoencoder predicted image (right), showing a resolution increase of around a factor of 5. Also shown is the catalogue generated ‘target’ image (centre), which has never been seen by the predicting model. The flux scale is in Jy/beam. A comparable deconvolution product can be created from the \texttt{XID+} prior-based deblending.}    \label{fig:deconvolution_comparison} \end{figure*}

The impressive deconvolution achieved from the hyperspectral imaging is not unique to the \texttt{XID+} algorithms. In this appendix we demonstrate this through an alternative method for prior catalogue generation, using a machine learning model trained to super-resolve the full range of PRIMA bands into a single output. We use a denoising autoencoder adapted from that developed for {\it Herschel} SPIRE $500\,\mu$m imaging by \citet{Lauritsen2021}. Unlike \texttt{XID+}, it does not assume the positions of a set of point sources extracted at shorter wavelengths as a Bayesian prior, but rather predicts the properties of the image from the training set of shorter wavelength data. The model was trained using cut-outs of the simulated hyperspectral PRIMAger imaging from this paper, with the target resolution for the longest wavelength data being that of the shortest wavelength imaging, i.e. a resolution improvement of a factor of approximately five. Figure \ref{fig:deconvolution_comparison} shows the results of this deconvolution in a segment of the simulated image; also shown is the PRIMA catalogue generated target image for the same region of sky. The model has never been exposed to this target image. A comprehensive analysis of the statistical properties of this alternative PRIMA deconvolution is deferred to a later paper, though it is clear that the deconvolution capacity is a general property of PRIMA's hyperspectral imaging, and not simply specific to prior-based deblending algorithms.  


\bsp	
\label{lastpage}
\end{document}